\documentclass[reqno,12pt]{amsart}

\usepackage[centertags]{amsmath}
\usepackage{amsfonts}
\usepackage{amssymb}
\usepackage{amsthm}
\usepackage{newlfont}
\usepackage{stmaryrd}
\usepackage{mathrsfs}
\usepackage{pstricks,pst-node}
\usepackage{palatino}  
\usepackage{fancyhdr}
\usepackage{graphicx}
\usepackage{fancybox}
\usepackage{multibox}
\usepackage{geometry}
\usepackage[dvips]{epsfig}
\usepackage{eepic}
\usepackage{pst-plot}

\theoremstyle{plain}
  \newtheorem{theorem}{Theorem}[section]   
  \newtheorem{corollary}[theorem]{Corollary}
  \newtheorem{proposition}[theorem]{Proposition}
  \newtheorem{lemma}[theorem]{Lemma}

  \newtheorem{definition}[theorem]{Definition}

  \newtheorem{remark}[theorem]{Remark}

 \DeclareMathOperator{\ext}{Ext}
\DeclareMathOperator{\Int}{Int} \DeclareMathOperator{\supp}{Supp}

\let\al=\alpha \let\be=\beta \let\de=\delta \let\ep=\epsilon
\let\ve=\varepsilon  \let\ga=\gamma 
 \let\la=\lambda \let\om=\omega 
\let\si=\sigma

\let\De=\Delta \let\Ga=\Gamma \let\La=\Lambda \let\Om=\Omega

\newcommand{\caB}{{\mathcal B}}
\newcommand{\caC}{{\mathcal C}}
\newcommand{\caD}{{\mathcal D}}

\newcommand{\caG}{{\mathcal G}}

\newcommand{\caK}{{\mathcal K}}

\newcommand{\caO}{{\mathcal O}}
\newcommand{\caP}{{\mathcal P}}

\newcommand{\caZ}{{\mathcal Z}}

\newcommand{\scD}{{\mathscr D}}

\newcommand{\bbC}{{\mathbb C}}

\newcommand{\bbE}{{\mathbb E}}

\newcommand{\bbN}{{\mathbb N}}

\newcommand{\bbR}{{\mathbb R}}

\newcommand{\bbZ}{{\mathbb Z}}

\newcommand{\opunit}{\text{1}\kern-0.22em\text{l}}
\newcommand{\funit}{\mathbf{1}}

\newcommand{\frB}{{\mathfrak B}}
\newcommand{\frC}{{\mathfrak C}}
\newcommand{\frD}{{\mathfrak D}}

\newcommand{\frZ}{{\mathfrak Z}}

\newcommand{\bsE}{{\boldsymbol E}}

\newcommand{\bsP}{{\boldsymbol P}}

\newcommand{\ie}{i.e.\;}

\newcommand{\un}[1]{\underline{#1}}

\newcommand{\rel}{\,|\,}

\newcommand{\pa}{_\bullet}

\newcommand{\pair}[1]{\langle{#1}\rangle}

\newcommand{\con}{_{\text{con}}}

\newcommand{\id}{\mathrm{d}}

\newcommand{\dL}{\partial\La}

 \DeclareMathOperator{\Ext}{Ext}

\DeclareMathOperator{\card}{card}
\DeclareMathOperator{\var}{Var}
\DeclareMathOperator{\cn}{Con}
\DeclareMathOperator{\dom}{Dom}

\newcommand{\A}{\ensuremath{\mathcal{A} }}

\newcommand{\bde}{\begin{definition}}
\newcommand{\ede}{\end{definition}}
\newcommand{\beq}{\begin{equation}}
\newcommand{\eeq}{\end{equation}}
\newcommand{\ben}{\begin{enumerate}}
\newcommand{\een}{\end{enumerate}}
\newcommand{\ble}{\begin{lemma}}
\newcommand{\ele}{\end{lemma}}
\newcommand{\bpr}{\begin{proof}}
\newcommand{\epr}{\end{proof}}
\newcommand{\bet}{\tilde\be}
\newcommand{\lra}{\leftrightarrow}

\newcommand{\inc}{\not\sim}

\renewcommand{\pa}{\partial}

\begin{document}

\title[Ising with random boundary condition]
  {On the Ising model with random boundary condition}

\author{A.~C.~D.~van Enter}%
\address{Institute for Theoretical Physics \\ University of Groningen \\
The Netherlands}%
\email{A.C.D.van.Enter@phys.rug.nl}%

\author{K.~Neto\v cn\'y}%
\address{Eurandom \\ Eindhoven \\ The Netherlands}%
\email{netocny@eurandom.tue.nl}%

\author{H.~G.~Schaap}%
\address{Institute for Theoretical Physics \\ University of Groningen \\
The Netherlands}%
\email{H.G.Schaap@phys.rug.nl}%





\begin{abstract}
The infinite-volume limit behavior of the 2d Ising model under
possibly strong random boundary conditions is studied. The model
exhibits chaotic size-dependence at low temperatures and we prove
that the `+' and `-' phases are the only almost sure limit Gibbs
measures, assuming that the limit is taken along a sparse enough
sequence of squares. In particular, we provide an argument to show
that in a sufficiently large volume a typical spin configuration
under a typical boundary condition contains no interfaces. In order
to exclude mixtures as possible limit points, a detailed multi-scale
contour analysis is performed.
\end{abstract}

\maketitle

\section{Introduction}

A fundamental problem in equilibrium statistical mechanics is to
determine the set of  physically accessible thermodynamic states for
models defined via a family of local interactions. Usually
\cite{EFS,Ge} one interprets the extremal elements of the set of
translationally invariant Gibbs measures as the pure thermodynamic
phases of the model. In particular this means that one gathers all
periodic or quasiperiodic extremal Gibbs measures into
symmetry-equivalent classes and identifies the latter with the pure
phases. Examples are the ferromagnetic, the antiferromagnetic,
crystalline or quasicrystalline phases exhibited by various models.
In this approach one does not consider either interface states or
mixtures as pure phases. The mixtures allow for a unique
decomposition into the extremal measures and are traditionally
interpreted in terms of a lack of the knowledge about the
thermodynamic state of the system. They can also be classified as
less stable than the extremal measures \cite{Haag,Narnhofer}. It is
thought that interface states which are extremal Gibbs measures are
more stable than mixed states, but less so than pure phases. However,
such an ``intrinsic'' characterization has not been developed. Note,
moreover, that in disordered systems such as spin glasses, the
stability of pure phases is a priori not clear and characterizing
them remains an open question.

An efficient strategy for models with a simple enough
structure of low-tempera{\-}ture phases is to associate these with suitable
\emph{coherent boundary conditions}. The latter are usually chosen as
\emph{ground states} of the model. As an example, the `+' and `-' Ising
phases can be obtained by fixing the constant `+', respectively the
constant `-' configurations at the boundaries and by letting the volume tend
to infinity. This idea has been generalized to a wide class
of models with both a finite and a `mildly' infinite number of
ground states, and
is usually referred to as the \emph{Pirogov-Sinai theory}
\cite{BI89,BS,Z84,PS1,PS2}.
The main assumption is that the different ground states are separated
by high enough energy barriers, which can be described in terms of domain walls,
referred to as contours. A useful criterion to check this so-called Peierls condition
is within the formalism of $m$-potentials due to Holzstynski and Slawny \cite{HS}.

An alternative strategy is to employ a boundary condition that does not
favor any of the phases. Examples are the free and periodic boundary conditions for
the zero-field
Ising model, or the periodic boundary conditions for the Potts model at
the critical temperature. In all these cases, an infinite-volume Gibbs
measure is obtained that is a homogenous mixture of all phases.

Another scenario has been expected to occur for spin glasses. Namely,
Newman and Stein have conjectured \cite{Ne97,NeSt92,NeSt97,NeSt02,NeSt03}
that  some spin glass models under symmetric boundary conditions exhibit
non-convergence to a single thermodynamic limit measure, a phenomenon called
\emph{chaotic size dependence} (see also \cite{FH,Lo,vE90}).
In this case, both the set of limit points of the sequence of the
finite-volume Gibbs measures and their empirical frequency along the
sequence of increasing volumes are of interest, and the formalism of \emph{metastates}  has been developed \cite{NeSt97,NeSt02,NeSt98} to deal with these phenomena.
These arguments have been made rigorous for a class of mean-field models \cite{BovGay,Kue97,Kue98,BovvENie,vES02,NS03,Kue98pr}, whereas no such results are
available for short-range spin glasses. For some general background on spin glasses and disordered models we refer to \cite{B,Fr,Ku,T1}.

A natural toy-problem where the usual contour methods can be used
in the regime of chaotic size-dependence is the zero field Ising model with
the boundary condition  sampled from a random distribution which is
symmetric under the spin flip. In dimension 2 or more and at
any subcritical temperature (including  $T = 0$) the finite-volume Gibbs measures are
expected to oscillate randomly between the `+' and the `-' phases,
demonstrating the chaotic size dependence with exactly two limit points
coinciding with the thermodynamic phases of the model \cite{NeSt92}.
In particular, one does not expect either any interface (e.g. Dobrushin)
Gibbs states or any non-trivial statistical mixtures to occur as the limit points.
This problem was addressed in \cite{EMN} where the conjecture was rigorously proven
as the almost sure picture in the regime of the weak boundary coupling. In this
regime, the boundary
bonds are made sufficiently weaker w.r.t.\ the bulk bonds so that the
interface configurations become damped exponentially with
the size of the system, uniformly for \emph{all} boundary conditions. Hence,
all translationally non-invariant Gibbs measures are forbidden as possible
limit points and one only needs to prove that the mixtures do not
appear with probability 1.

In this paper we continue this study by removing the weakness
assumption on the boundary bonds. To be specific, we consider the 2d
Ising model with the random boundary condition sampled from the
symmetric i.i.d.\ field $\{-1,1\}^{\bbZ^2}$ and coupled to the system
via the \emph{bulk} coupling constant. The conjecture remains true in
this case and the crucial novelty of our approach is a detailed
multi-scale analysis of contour models in the regime where
realizations of the boundary condition are allowed that violate the
`diluteness' (Peierls) condition, possibly making interfaces likely
configurations. To be precise, these interfaces can have large Gibbs
probabilities for certain boundary conditions, but we will show that
such boundary conditions are sufficiently unlikely to occur for large
volumes. An important side-result is the almost sure absence of
interface configurations. This means that for a typical boundary
condition, the probability of the set of configurations containing an
interface tends to zero in the infinite-volume limit. Note that this
excludes interfaces in a stronger way than the familiar result about
the absence of translationally non-invariant Gibbs measures in the 2d
Ising model \cite{Ga,Aiz,Hi}.  Indeed, the absence of fluctuating
interfaces basically means that not only the expectations of local
functions but also their space averages (e.g.\ the volume-averaged
magnetization) have only two limit points, corresponding to the two
Ising phases. Hence, we believe that our techniques allow for a
natural generalization to any dimension $d\geq 2$. However, as
already argued in \cite{EMN}, in dimensions $d \geq 4$, the set
$\{\mu^+,\mu^-\}$ is expected (and partially proven) to be the \emph{almost sure} set of limit measures, the limit being taken along the regular sequence of cubes. On the other hand, for $d = 2,3$ the same result can only be obtained if the limit is taken along a sparse enough sequence of cubes. In the latter case it remains an open problem to analyze the set of limit points along the regular sequence of cubes. Our conjecture is that
the almost sure set of limit points coincides then with the set of all translationally invariant Gibbs measures, i.e.\ including the mixtures.

The structure of the paper is as follows. We will first introduce our notation in section 2, and describe our results in section 3. Then in sections 4 and 5 we will introduce a contour representation of the model and set up our cluster expansion formalism. In section 6 we first exclude the occurrence of interfaces. In the rest of the paper we develop a multiscale argument, providing a weak version of the local limit theorem to show that no mixed states can occur as limit points in the infinite-volume limit. Two general results, the first one on
a variant of the cluster expansion convergence criteria for polymer models and the second one on local limit upper bounds, are collected in two Appendices.

\section{Set-up}\label{sec: set-up}

We consider the two-dimensional square lattice $\bbZ^2$ and use
the symbols $\si,\eta,\ldots$ for the maps
$\bbZ^2\mapsto\{-1,1\}$. They are called \emph{spin configurations}
and the set of all spin configurations is
$\Om = \{-1,1\}^{\bbZ^2}$. Furthermore, the symbol $\si_A$ is used for
the restriction of a spin configuration $\si\in\Om$ to the set
$A\subset\bbZ^2$. If $A = \{x\}$, we write $\si_x$ instead.
The set of all restrictions of $\Om$ to the set $A$ is $\Om_A$.
\\ A function $f:\Om\mapsto\bbR$ is called \emph{local} whenever
there is a finite set $D\subset\bbZ^2$ such that
$\si_{D} = \si'_{D}$ implies $f(\si) = f(\si')$. The smallest
set with this property is called the \emph{dependence set} of the
function $f$ and we use the symbol $\caD_f$ for it. To every local
function $f$ we assign the supremum norm
$\|f\| = \sup_{\si\in\Om}|f(\si)|$.
\\ The spin configuration space $\Om$ comes equipped with the
product topology, which is followed by the weak topology on the
space $M(\Om)$ of all probability measures on $\Om$. The latter is
introduced via the collection of seminorms
\begin{equation}
  \|\mu\|_X = \sup_{\|f\| = 1 \atop \caD_f \subset X} |\mu(f)|
\end{equation}
upon all finite $X\subset\bbZ^2$. Then, the weak topology is
generated by the collection of open balls
$B_X^\ep(\mu) = \{\nu;\ \|\nu-\mu\|_X < \ep\}$,
$\ep > 0$, $X$ finite, and a sequence
$\mu_n \in M(\Om)$ weakly converges to $\mu$ if and only if
$\|\mu_n - \mu\|_X \to 0$ for all finite $X\subset\bbZ^2$.
Under the weak topology, $M(\Om)$ is compact.

We consider a collection of the Hamiltonians
$H_\La^\eta: \Om_\La \mapsto \bbR$ for all square volumes
$\La = \La(N)$, $N = 1,2,\ldots$,
\begin{equation}
  \La(N) = \{x\in\bbZ^2;\, \|x\|\leq N\}  \qquad
  \|x\| = \max\{|x_1|,|x_2|\}
\end{equation}
and \emph{boundary conditions} $\eta\in\Om$. The Hamiltonians are
given by
\begin{equation}\label{eq: ham}
  H_\La^\eta(\si_\La) = -\be\sum_{\langle x,y \rangle \subset \La}
  (\si_x \si_y - 1) - \be\sum_{\langle x,y \rangle \atop x \in \La,\, y \in \La^c}
  \si_x \eta_y
\end{equation}
where $\langle x,y \rangle$ stands for pairs of nearest neighboring
sites, \ie such that
$\|x-y\|_1 :=
|x_1 - y_1| + |x_2 - y_2| = 1$, and $\La^c = \bbZ^2 \setminus \La$.
We consider the ferromagnetic case, $\be > 0$. Following a familiar
framework, we introduce the
\emph{finite-volume Gibbs measure} $\mu_\La^\eta \in M(\Om)$ by
\begin{equation}\label{eq: Gibbs measure}
  \mu_\La^\eta(\si) = \frac{1}{Z_\La^\eta} \exp[-H_{\La}^{\eta}(\si_\La)]\,
  \opunit_{\{\si_{\La^c} = \eta_{\La^c}\}}
\end{equation}
and define the set $\caG_\be$ of (infinite-volume) \emph{Gibbs
measures}, $\caG_\be$, as the weak closure of the convex hull over
the set of all weak limit points of the sequences
$(\mu_{\La(N)}^\eta)_{N\rightarrow\infty}$, $\eta \in \Om$.
A standard result reads that there exists $\be_c$ such that for
any $\be > \be_c$ the set of Gibbs measures
$\caG_\be = \{\al\mu^+ + (1-\al)\mu^-;\, 0 \leq \al \leq 1\}$. Here,
the extremal measures $\mu^\pm$ are translation-invariant, they
satisfy the symmetry relation
$\int \id\mu^+(\si)\,f(\si) = \int\id\mu^-(\si)\,f(-\si)$,
and can be obtained as the weak limits
$\lim_{N\rightarrow\infty} \mu_{\La(N)}^\eta$ for $\eta \equiv \pm 1$.

\section{Results}
We consider the limit behavior of the sequence of
finite-volume Gibbs measures ($\mu_{\La(N)}^\eta)_{N\in\bbN}$
under boundary conditions $\eta$ sampled from the i.i.d.\
symmetric random field
\begin{equation}
  \bsP\{\eta_x = 1\} = \bsP\{\eta_x = -1\} = \frac{1}{2}
\end{equation}
Our first result concerns the almost sure structure of the set of
all limit points of the sequence of the finite-volume Gibbs
measures, the limit being taken along a sparse enough sequence of
squares.
\begin{theorem}\label{thm: main}
For arbitrary $\om > 0$ there is a $\be_1 = \be_1(\om)$ such that
for any $\be \geq \be_1$ the set of all weak limit points of any
sequence
$(\mu_{\La(k_N)})_{N=1,2,\ldots}$, $k_N \geq N^{2+\om}$,
is $\{\mu^+,\mu^-\}$, $\bsP$-a.s.
\end{theorem}
\begin{remark}
The above theorem does not exclude other measures as the almost
sure limit points, provided that other (non-sparse) sequences of
squares are taken instead. Actually, our conjecture is that, for
$\be$ large enough, the set of all weak limit points of
$(\mu_{\La(N)})_{N=1,2,\ldots}$ coincides $\bsP$-a.s.\ with
$\caG_\be$. On the other hand, in dimension $3$, it is rather expected
to coincide with the set of all translation-invariant Gibbs
measures, and, in any dimension higher than $3$, with the set
$\{\mu^+,\mu^-\}$.
\end{remark}
\begin{remark}
A modification of the Hamiltonian \eqref{eq: ham} is obtained by
re-scaling the boundary coupling by a factor $\la$ to get
\begin{equation}
  H_\La^{\la,\eta}(\si_\La) = -\be\sum_{\langle x,y \rangle \subset \La}
  (\si_x \si_y - 1) - \la \be\sum_{\langle x,y \rangle \atop x \in \La,\, y \in \La^c}
  \si_x \eta_y
\end{equation}
In this case, the claim of Theorem~\ref{thm: main} for the
sequence of the finite-volume Gibbs measures
\begin{equation}
  \mu_\La^{\la,\eta}(\si) = \frac{1}{\caZ_\La^{\la,\eta}}
  \exp[-H_{\La}^{\la,\eta}(\si_\La)]\,
  \opunit_{\{\si_{\La^c} = \eta_{\La^c}\}}
\end{equation}
was proven in \cite{EMN} under the condition that
$|\la|$ is small enough (= the boundary coupling is sufficiently
weak w.r.t.\ the bulk one). It was also shown that
$\{\mu^+,\mu^-\}$ is the almost sure set of limit points of the
sequence $(\mu_{\La(N)}^\eta)_{N\in\bbN}$, provided that the space
dimension is at least $4$.
\end{remark}
To reveal the nature of all possible limit points that can appear
along the sequence of squares $\La(N)$, $N = 1,2,\ldots$, we study
the empirical frequency for the finite-volume Gibbs states from
the sequence $(\mu_{\La(N)}^\eta)_{N\in\bbN}$ to occur in a fixed
set of measures. More precisely, for any set $B \subset M(\Om)$,
boundary condition $\eta \in \Om$, and
$N =1,2,\ldots$, we define
\begin{equation}
  Q_N^{B,\eta} = \frac{1}{N} \sum_{k=1}^{N}
  \opunit_{\{\mu_{\La(k)}^\eta \in B\}}
\end{equation}
The next theorem shows the \emph{null-recurrent} character of all
measures different from both $\mu^+$ and $\mu^-$. We use the
notation $\bar B$ and $B^0$ for the weak closure and the weak
interior of $B$, respectively.
\begin{theorem}\label{thm: null-recurrence}
There is $\be_2$ such that for any $\be \geq \be_2$ and any set
$B \subset M(\Om)$, one has
\begin{equation}
  \lim_{N\uparrow\infty} Q_N^{B,\eta} =
  \begin{cases}
    0  &  \text{if } \mu^+,\mu^- \not\in \bar B
  \\
    \frac{1}{2}  &  \text{if } \mu^\pm \in B^0 \text{ and }
    \mu^\mp \not\in \bar B
  \\
    1  &  \text{if } \mu^+,\mu^- \in B^0
  \end{cases}
\end{equation}
with $\bsP$-probability $1$.
\end{theorem}
Both theorems follow in a straightforward way from the following key
estimate that will be proven in the sequel of the paper.

\begin{proposition}\label{prop: basic-est}
Given $\al > 0$, there is a $\be_0 = \be_0(\al)$ such that for any
$\be \geq \be_0$, $\ep > 0$ and $X \subset \bbZ^d$ finite,
\begin{equation}
  \varlimsup_{N\to\infty} N^{\frac{1}{2} - \al}\,
  \bsP\{(\|\mu_{\La(N)}^\eta - \mu^+\|_X \wedge
  \|\mu_{\La(N)}^\eta - \mu^-\|_X) \geq \ep\} < \infty
\end{equation}
\end{proposition}
\begin{remark}
The proposition claims that, for a \emph{typical} $\eta \in \Om$,
the finite-volume Gibbs measures are expected to be near the
extremal Gibbs measures $\mu^\pm$. The above probability
upper-bound of the form $\caO\bigl( N^{-\frac{1}{2} + \al} \bigr)$
will be proven by means of a variant of the local limit theorem
for the sum of weakly dependent random variables. Although we conjecture
the correct asymptotics to be of order $N^{-\frac{1}{2}}$, the proof of
\emph{any} lower bound goes beyond
the presented technique. This is why the detailed
structure of the almost sure set of the limit Gibbs measures is
not available, except for the limits taken along sparse enough
sequences of squares.
\end{remark}
\begin{proof}[Proof of Theorem~\ref{thm: main}]
Given $\om > 0$, we choose an $\al < \om / (2(2+\om))$ and define
$\be_1(\om) = \be_0(\al)$. Let $\be \geq \be_1(\om)$ and
$k_N \geq N^{2+\om}$.

First let $\mu \not\in \{\mu^+,\mu^-\}$. There exists a weakly
open set $B \subset M(\Om)$ such that $\mu \in B$ and
$\mu^+,\mu^- \not\in \bar B$. Choosing
a finite set $X \subset \bbZ^2$ and $\ep > 0$ such that
$B_X^\ep(\mu^\pm) \cap B = \emptyset$,
Proposition~\ref{prop: basic-est} gives the bound
\begin{equation}
\begin{split}
  &\bsP\{\mu_{\La(k_N)}^\eta \in B\}
  \leq \bsP\{\mu_{\La(k_N)}^\eta \not\in
  B_X^\ep(\mu^+) \cup B_X^\ep(\mu^-)\}
\\
  &= \caO\large( k(N)^{-\frac{1}{2} + \al} \large)
  = \caO\large( N^{-1+\al(2+\om) - \frac{\om}{2}} \large)
\end{split}
\end{equation}
Since $\sum_N \bsP\{\mu_{\La(k_N)}^\eta \in B\} < \infty$, the set
$B$ contains $\bsP$-a.s.\ no limit points of the
sequence $\mu_{\La(k_N)}^\eta$ due to the Borel-Cantelli argument.
Hence, with $\bsP$-probability 1, $\mu$ is not a limit point.

To prove that both $\mu^+$ and $\mu^-$ are $\bsP$-a.s.\ limit points, take
any finite set of sites $X$ and $\ep > 0$ such that
$B_X^\ep(\mu^+) \cap B_X^\ep(\mu^-) = \emptyset$. By the symmetry
of the distribution,
$\bsP\{\mu_{\La(k_N)} \in B_X^\ep(\mu^+)\}
= \bsP\{\mu_{\La(k_N)} \in B_X^\ep(\mu^-)\}$ and, employing
Proposition~\ref{prop: basic-est} again,
$\lim_N \bsP\{\mu_{\La(k_N)} \in B_X^\ep(\mu^\pm)\} = \frac{1}{2}$.
By the Borel-Cantelli and the compactness arguments, the weak
closure $\bar B_X^\ep(\mu^\pm)$ contains a limit point,
$\bsP$-a.s. As $\mu^\pm = \cap_{X,\ep} \bar B_X^\ep(\mu^\pm)$, the
statement is proven.
\end{proof}

\begin{proof}[Proof of Theorem~\ref{thm: null-recurrence}]
Choose $\be_2 = \be_0(\al)$ for an arbitrary
$\al \in (0,\frac{1}{2})$ and assume $\be \geq \be_2$,
$B \in M(\Om)$. Using the notation
$q_N^{B,\eta} = \bsP{\{\mu_{\La(N)}^\eta \in B\}}$ and
repeating the reasoning in the proof of Theorem~\ref{thm: main},
one gets
\begin{equation}
\begin{split}
  \bsE\,\opunit_{\{\mu_{\La(N)}^\eta \in B\}} &= q_N^{B,\eta}
\\
  &=
  \begin{cases}
    \caO\large(N^{-\frac{1}{2}+\al}\large) \rightarrow 0
    & \text{if } \mu^+,\mu^- \not\in \bar B
  \\
    \frac{1}{2} - \caO\large(N^{-\frac{1}{2}+\al}\large) \rightarrow
    \frac{1}{2} &
    \text{if } \mu^\pm \in B^0 \text{ and } \mu^\mp \not\in \bar B
  \\
    1 - \caO\large(N^{-\frac{1}{2}+\al}\large) \rightarrow 1 &
    \text{if } \mu^\pm \in B^0
  \end{cases}
\end{split}
\end{equation}
and
\begin{equation}
  \var\,\opunit_{\{\mu_{\La(N)}^\eta \in B\}}
  = q_N^{B,\eta}(1 - q_N^{B,\eta}) \leq \frac{1}{4}
\end{equation}
Hence,
$\sum_N \frac{1}{N^2}
\var\,\opunit_{\{\mu_{\La(N)}^\eta \in B\}} < \infty$ and since
the functions $\opunit_{\{\mu_{\La(N)}^\eta \in B\}}$,
$N=1,2,\ldots$ are independent, the result immediately follows
from the strong law of large numbers \cite{Du}.
\end{proof}

\section{Geometrical representation of the model}\label{sec: geometrical}

We define the \emph{dual lattice} $(\bbZ^2)^* = \bbZ^2 + (1/2,1/2)$.
The (unordered) pairs of nearest neighboring sites
$\langle x,y \rangle \subset \bbZ^2$ are called
\emph{bonds} and to every bond we assign a unique \emph{dual bond}
$\langle x^*,y^* \rangle \equiv \langle x,y \rangle^* \subset (\bbZ^2)^*$.
Given a set of dual bonds $A^*$, we use the symbol
$|A^*|$ to denote the number of all dual bonds in $A^*$. Further,
with a slight abuse of notation, we also write $x^* \in A^*$ whenever
there exists a dual bond $\langle x^*,y^* \rangle \in A^*$, i.e.\
$A^*$ also stands for the corresponding set of dual sites.

Any set $A^*$ of dual bonds is called
\emph{connected} whenever for any dual sites
$x^*,y^* \in A^*$ there exists a sequence of dual bonds
$\langle x^*,x^*_1 \rangle, \langle x^*_1,x^*_2 \rangle,
\ldots, \langle x^*_{k-1},y^* \rangle \in A^*$.
The distance $d[A^*,B^*]$ of the sets of dual bonds
$A^*,B^*$ is defined as the smallest integer $k$ such that
there exist
$x^* \in A^*$, $y^* \in B^*$, and a sequence of dual bonds
$\langle x^*,x^*_1 \rangle, \langle x^*_1, x^*_2
\rangle, \ldots, \langle x^*_{k-1},y^* \rangle \subset (\bbZ^2)^*$.
Similarly, a set of sites $A \subset \bbZ^2$ is called
\emph{connected} whenever for all $x,y \in A$ there exists a sequence
of bonds $\langle x,x_1 \rangle, \langle x_1,x_2 \rangle, \ldots,
\linebreak \langle x_{k-1},y \rangle \subset A$.  Correspondingly, the distance $d[A,B]$
of the sets $A,B \subset \bbZ^2$ is understood in the sense of the
$\|.\|_1$-norm.

In the sequel we assume that a volume $\La = \La(N)$ is fixed and we
define the \emph{boundary} $\partial\La$ as the set of all dual bonds
$\langle x,y \rangle^*$ such that $x \in \La$ and $y \in \La^c$.
In general, $\partial A$, $A \subset \La$ is the set of all dual
bonds $\langle x,y \rangle^*$, $x \in A$, $y \in \La^c$. For any
subset $P \subset \partial\La$ we use the symbol $\un{P}$ to denote
the set of all sites $y \in \La^c$ such that there is a (unique) bond
$\langle x,y \rangle^* \in P$, $x \in  \La$.
If $\un{P}$ is a connected set of sites, then $P$ is called a
\emph{boundary interval}. Obviously, any boundary interval is a
connected set of dual bonds, however, the opposite is not true.
However, any set $P \subset \partial\La$ has a unique decomposition
into a family of (maximal) boundary intervals. Furthermore, consider
all connected sets $P_i$ of dual bonds satisfying
$P \subset P_i \subset \pa\La$ which are minimal in the sense of
inclusion. The smallest of these sets is called
$\cn(P)$ (in the case of an equal size take the first one in the
lexicographic order) and we use the shorthand
$|P|\con = |\cn(P)|$.
Finally, we define the \emph{corners} of $\La(N)$ as the dual sites
$x^*_{C,1} = (-N-1/2,-N-1/2)$, $x^*_{C,2} = (N+1/2,-N-1/2)$,
$x^*_{C,3} = (N+1/2,N+1/2)$, and $x^*_{C,4} = (-N-1/2,N+1/2)$.

\subsection*{Pre-contours}

\begin{figure}
\begin{center}
\includegraphics[scale=0.6]{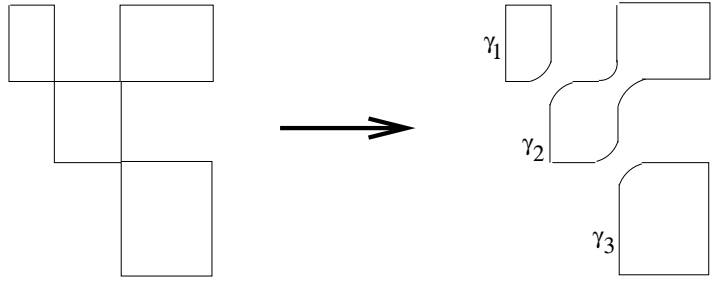}
\caption{Pre-contours constructed via the rounding corner procedure.}
\end{center}
\end{figure}

Given a configuration $\si \in \Om_\La = \{-1,+1\}^\La$, the dual
bond $\langle x,y \rangle^*$ to a bond
$\langle x,y \rangle \subset \La$ is called \emph{broken} whenever
$\si_x \neq \si_y$, and the set of all the broken dual bonds is denoted by $\De_\La(\si)$.
In order to define a suitable decomposition of the set
$\De_\La(\si)$ into components, we take advantage of a certain freedom
in such a construction to obtain the components with suitable
geometrical properties. In this first step, we define the
\emph{pre-contours} as follows. Consider all maximal connected
components of the set of dual bonds $\De_\La(\si)$. By the standard
`rounding-corner' procedure, see Figure~1, we further split them into
connected (not necessarily disjoint) subsets,
$\ga$, which can be identified with (open or closed) simple curves.
Namely,
\[
  \ga = \{\langle x^*_0,x^*_1 \rangle, \langle x^*_1,x^*_2 \rangle,
  \ldots, \langle x^*_{k-1},x^*_k \rangle\}
  \qquad k \in \bbN
\]
such that if $x^*_i = x^*_j$, $i \neq j$, then $\{i,j\} = \{0,k\}$
and $\ga$ is closed. Otherwise, $x^*_i \neq x^*_j$ for all $i \neq j$
and $\ga$ is open with $x^*_0, x^*_k \in \pa\La$.

These $\ga$ are called \emph{pre-contours} and we use the symbol
$\tilde\caD_\La(\si)$ for the set of all pre-contours corresponding
to $\si$; write also $\tilde\scD_\La =
\{\tilde\caD_\La(\si),\,\si \in \Om_\La\}$ and use the symbol
$\tilde\caK_\La$ for the set of all pre-contours in $\La$.
Any pair of pre-contours
$\ga_1,\ga_2 \in \tilde\caK_\La$ is called \emph{compatible} whenever
there is a configuration $\si \in \Om_\La$ such that
$\ga_1,\ga_2 \in \tilde\caD_\La(\si)$. A set of pairwise compatible
pre-contours is called a \emph{compatible set}. Obviously,
$\tilde\scD_\La$ is simply
the collection of all compatible sets of pre-contours from
$\tilde\caK_\La$. Intuitively, the pre-contours that are closed
curves coincide with the familiar Ising contours, whereas the
pre-contours touching the boundary become open curves.

\begin{figure}
\begin{center}
\includegraphics[scale=0.6]{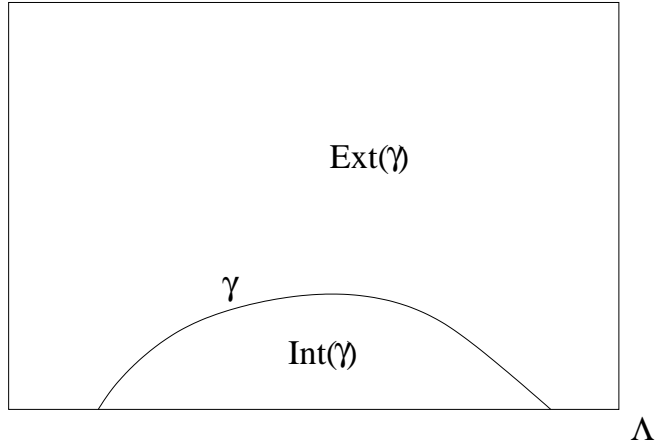}
\caption{Small boundary pre-contour.}
\end{center}
\end{figure}

Obviously, $\Om_\La \mapsto \tilde\scD_\La$ is a two-to-one map with
the images of the configurations $\si$ and $-\si$ being identical. In
order to further analyze this map, we introduce the concept of
interior and exterior of the pre-contours briefly as follows (the
details can be found in \cite{BK95,EMN}). If $\si \in \Om_\La$ is a
configuration such that $\tilde\caD_\La(\si) = \{\ga\}$, then there
is a unique decomposition of the set $\La$ into a pair of disjoint
connected subsets,
$\La = \La_1 \cup \La_2$, such that for any bond
$\langle x,y \rangle$, $x \in \La_1$, $y \in \La_2$, one has
$\langle x,y \rangle^* \in \ga$.
These are called the \emph{exterior}, $\Ext(\ga)$, and the
\emph{interior}, $\Int(\ga)$, where the assignment is given by the
following procedure. We distinguish three mutually exclusive classes
of pre-contours:
\begin{enumerate}
\item[i)]
\emph{Bulk pre-contours.}\\
$\dL = \dL_1$. Then, $\Ext(\ga) := \La_1$ and $\Int(\ga) := \La_2$,
see Figure~1.
\item[ii)]
\emph{Small boundary pre-contours.}\\
$\La_1$ contains at least three corners of $\La$ and
$\dL_2 \neq \emptyset$. Then,
$\Ext(\ga) := \La_1$ and $\Int(\ga) := \La_2$,
see Figure~2.
\item[iii)]
\emph{Interfaces.}\\
Both $\La_1$ and $\La_2$ contain exactly two corners of $\La$ and a)
$|\La_1| > |\La_2|$, or b) $|\La_1| = |\La_2|$ and $x_{C,1} \in
\La_1$. Then, $\Ext(\ga) := \La_1$ and $\Int(\ga) := \La_2$, see
Figure~3.
\end{enumerate}
The set $\pa\ga:=\pa\Int(\ga)$ is called the \emph{boundary} of the pre-contour $\ga$.

\begin{figure}
\begin{center}
\includegraphics[scale=0.6]{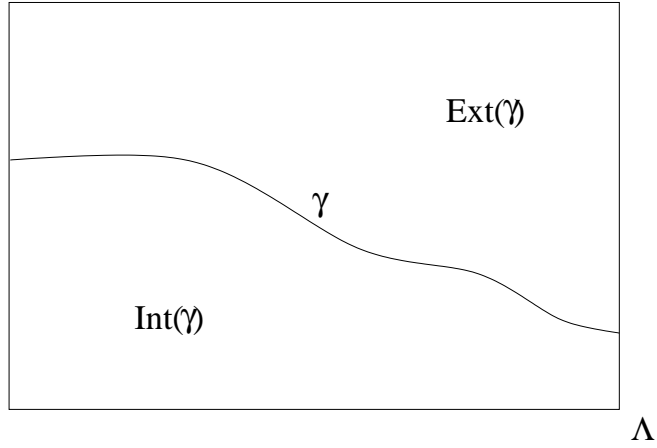}
\caption{Interface.}
\end{center}
\end{figure}

\subsection*{Contours}

Next, we define contours by gluing some boundary pre-contours together via the following
procedure. Any compatible pair of pre-contours $\ga_1,\ga_2 \in \tilde\caK_\La$ is called
\emph{boundary-matching} iff $\pa\ga_1 \cap \pa\ga_2 \neq \emptyset$. Any
compatible set of pre-contours such that the graph on this set
obtained by connecting the pairs of boundary-matching pre-contours
becomes connected is called a \emph{contour}. In particular, every
bulk pre-contour is boundary-matching with no other compatible
pre-contour. Therefore, every bulk pre-contour is trivially a
contour. We use the symbol
$\caD_\La(\si)$ for the set of all contours corresponding to
$\si\in\Om_\La$ and $\caK_\La$ for the set of all contours in $\La$.
Any pair of contours $\Ga_1,\Ga_2$ is compatible,
$\Ga_1 \sim \Ga_2$, whenever all pairs of pre-contours
$\ga_1 \in \Ga_1$, $\ga_2 \in \Ga_2$ are compatible, and we write
$\scD_\La$ for the set of all families of pairwise compatible
contours in $\La$. All the above geometrical notions naturally carry
over to contours and we define the exterior, $\ext(\Ga) :=
\cap_{\ga\in\Ga} \ext(\ga)$, the interior,
$\Int(\Ga) := \La \setminus \ext(\Ga)$ (in general, not a connected set anymore),
the boundary $\pa\Ga := \cup_{\ga\in\Ga} \pa\ga$, and the length
$|\Ga| := \sum_{\ga\in\Ga} |\ga|$. Similarly, if $\pa \in \scD_{\La}$
is a configuration of contours, let $\ext(\pa) := \cap_{\Ga\in\pa}
\ext(\Ga)$, $\Int(\pa) := \La \setminus \ext(\pa)$, and
$|\pa| := \sum_{\Ga \in \pa} |\Ga|$.

\begin{figure}
\begin{center}
\includegraphics[scale=0.6]{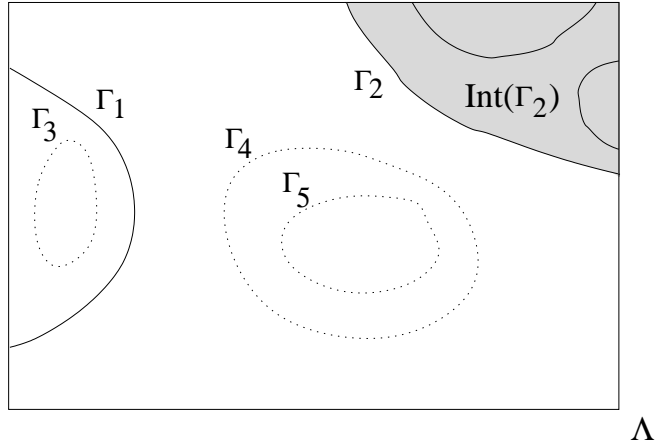}
\caption{Bulk and small boundary contours.}
\end{center}
\end{figure}

Eventually we arrive at the following picture. The set $\caK_\La$
of contours is a union of three disjoint sets of contours, namely
of the sets of all
\begin{enumerate}
\item[i)]
\emph{bulk (pre-)contours}.
\item[ii)]
\emph{small boundary contours} $\Ga$ defined by 1) $\pa\Ga \neq \emptyset$, and 2)
no pre-contour $\ga\in\Ga$ is an interface.
\begin{enumerate}
\item[a)]
\emph{simple} small boundary contours: the boundary $\pa\Ga$ contains no corner, i.e.\,
$\pa\Ga$ is a boundary interval.
\item[b)]
\emph{corner} small boundary contours: there is exactly one corner
$x^*_{C,i} \in \pa\Ga$.
\end{enumerate}
\item[iii)]
\emph{large boundary contours} $\Ga$, i.e.\ containing at least one interface
$\ga\in\Ga$.
\end{enumerate}
Examples of the bulk, small boundary, and large boundary contours are
given in Figures~4-5.

\begin{figure}
\begin{center}
\includegraphics[scale=0.6]{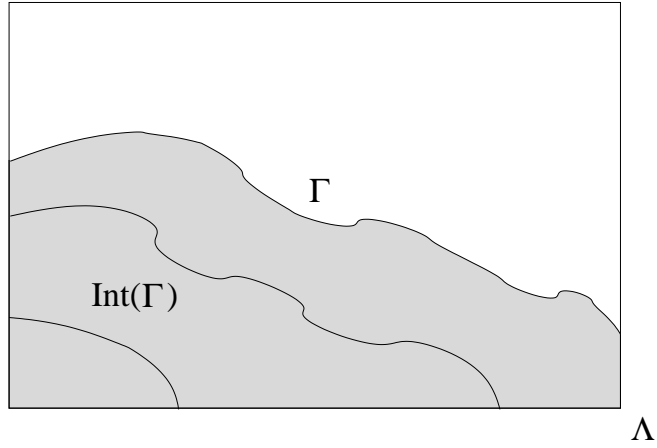}
\caption{Large boundary contour.}
\end{center}
\end{figure}

Furthermore, $\caD_\La(\si)$ is a two-to-one map
$\Om_\La\mapsto\scD_\La$ satisfying the spin-flip symmetry
$\caD_\La(\si) = \caD_\La(-\si)$. Since $\si$ takes a unique spin
value in the set $\ext(\caD_\La(\si))$, there is a natural
decomposition
$\Om_\La = \Om_\La^+ \cup \Om_\La^-$ according to this value, i.e.\
\begin{equation}
  \Om_\La^\pm := \{\si \in \Om_\La;\, \si|_{\ext(\caD_\La(\si))} = \pm 1\} = -\Om_\La^\mp
\end{equation}
As a consequence, $\caD_\La$ splits into a conjugated (by spin-flip
symmetry) pair of one-to-one maps $\Om_\La^\pm \mapsto \scD_\La$.
This enables us to represent the finite-volume Gibbs measure
\eqref{eq: Gibbs measure} in the form of a convex combination of two
conjugated constrained Gibbs measures as follows:
\begin{equation}\label{eq: representation-Gibbs}
\begin{split}
  \mu_\La^\eta(\si)
  &= \Bigl[ 1 + \frac{\caZ_\La^{-,\eta}}{\caZ_\La^{+,\eta}}
  \Bigr]^{-1} \nu_\La^{+,\eta}(\si)
  + \Bigl[ 1 + \frac{\caZ_\La^{+,\eta}}{\caZ_\La^{-,\eta}}
  \Bigr]^{-1}\nu_\La^{-,\eta}(\si)
\end{split}
\end{equation}
where we have introduced the Gibbs measure constrained to $\Om_\La^\pm$ by
\begin{equation}
  \nu_\La^{\pm,\eta}(\si) = \frac{1}{\caZ_\La^{\pm,\eta}}
  \exp[-H_{\La}^{\eta}(\si)]\, \funit_{\{\si \in \Om_\La^\pm\}}
\end{equation}
Moreover, for any $\si\in\Om_\La^\pm$, the Hamiltonian can be written as
\begin{equation}
  H_\La^\eta(\si) = E_\La^{\pm,\eta}(\partial) + 2\be \sum_{\Ga \in \partial} |\Ga|
\end{equation}
with $\pa = \caD_\La(\si)$, and we have introduced
\begin{equation}
  E_\La^{\pm,\eta}(\partial) =
  -\be\sum_{\langle x,y \rangle \atop x \in \La,\, y \in \La^c} \si_x \eta_y
\end{equation}
Finally, $\caZ_\La^{\pm,\eta}$ is essentially the partition function of a polymer model
\cite{KoPr}, see also Appendix~\ref{ap: cluster models},
\begin{align}\label{eq: polymer model}
  \caZ_\La^{\pm,\eta} &=
  \exp{(-E_\La^{\pm,\eta}(\emptyset))}\sum_{\partial\in\scD_{\La}}
  \prod_{\Ga\in\pa} \rho^{\pm,\eta}(\Ga)
\\ \intertext{where the polymers coincide with the contours and the polymer weights are defined by}
\label{eq: rhoGa}
  \rho^{\pm,\eta}(\Ga) &= \exp{\left(-2\be|\Ga|\right)}
  \exp{\left(-E^{\pm,\eta}(\Ga)+E^{\pm,\eta}(\emptyset)\right)}
\end{align}
By the spin-flip symmetry, we can confine ourselves to the `+' case
and use the shorthand notations $\rho^\eta(\Ga) := \rho^{+,\eta}(\Ga) =
\rho^{-,-\eta}(\Ga)$, $\caZ_\La^\eta := \caZ_\La^{+,\eta} =
\caZ_\La^{-,-\eta}$, $E_\La^\eta := E_\La^{+,\eta} =
E_\La^{-,-\eta}$, and $\nu_\La^\eta(\si) := \nu_\La^{+,\eta}(\si)
= \nu_\La^{-,-\eta}(-\si)$.
Moreover, the boundary $\pa\Ga$ of a contour $\Ga$ has a natural decomposition into components as follows. Let $\si \in \Om_\La^+$ be such that $\caD_\La(\si) = \{\Ga\}$. Then the `$\pm$' boundary component $\pa\Ga^\pm$ is defined as the set of all dual bonds
$\langle x,y \rangle^*$ such that $x \in \La$, $y \in \La^c$, $\si_x = \pm1$. With this definition, the contour weight \eqref{eq: rhoGa} is
\begin{equation}\label{eq: rhoGa1}
  \rho^\eta(\Ga) = \exp\Bigl[-2\be\bigl(|\Ga|
  + \sum_{x \in \un{\pa\Ga^-}}\eta_x\bigr)\Bigr]
\end{equation}

Using the representation~\eqref{eq: representation-Gibbs} of the finite-volume Gibbs measure $\mu_\La^\eta$, the strategy of our proof consists of two main parts:
\begin{enumerate}
\item
To prove that the constrained (random) Gibbs measure $\nu_\La^{\eta}$ asymptotically coincides with the Ising `+' phase, for almost all $\eta$.
\item
To show that a sufficiently sparse subsequence of the sequence of random free energy differences $\log \caZ_\La^{\eta} - \log \caZ_\La^{-\eta}$ has $+\infty$ and $-\infty$ as
the only limit points, for almost all $\eta$.
\end{enumerate}
Then, Proposition~\ref{prop: basic-est} follows almost immediately.
Moreover, we will show that for a $\bsP$-typical boundary condition
$\eta$ and a $\mu_\La^\eta$-typical configuration $\si \in \Om_\La$,
the corresponding set of pre-contours $\tilde\caD_\La(\si)$ contains
no interfaces.
\begin{theorem}\label{thm: interfaces}
There is $\be_3$ such that for any $\be \geq \be_3$ one has
\begin{equation}
  \lim_{N\uparrow\infty} \mu_{\La(N)}^\eta\{\tilde\caD_{\La(N)}(\si)
  \text{ contains an interface}\} = 0
\end{equation}
for $\bsP$-a.e.\ $\eta \in \Om$.
\end{theorem}
\begin{remark}
Note that the low-temperature result by Gallavotti~\cite{Ga},
extended to all subcritical temperatures in \cite{Aiz,Hi}, about the
absence of translationally non-invariant Gibbs measures in the 2d
Ising model does not exclude fluctuating interfaces under a suitably
arranged (`Dobrushin-like') boundary condition.  On the other hand,
the above theorem claims that a \emph{typical} boundary condition
gives rise to a Gibbs measure in which interfaces
\emph{anywhere} are suppressed. We mention this side-result to
demonstrate the robustness of the presented multi-scale approach and
to argue that it is essentially dimension-independent, the $d=2$ case
being chosen only for simplicity.
\end{remark}

It is easy to realize that, for a typical $\eta$, the polymer model~\eqref{eq: polymer model} fails the `diluteness' condition on the sufficient exponential decay of the polymer weights, which means one cannot directly apply the familiar formalism of cluster expansions. These violations of the diluteness condition occur locally along the boundary with low probability, and hence have typically low densities. Nevertheless, their presence on all scales forces a sequential, multi-scale, treatment.
Multi-scale methods have been employed at various occasions, such as for one-phase models in the presence of Griffiths singularities or for the random field Ising model \cite{BK1,BK2,Klein,FFS,FI,I}. In contrast to the usual case of cluster expansions one does not obtain analyticity (which
may not even be valid). In our approach, we loosely follow the ideas of Fr\"ohlich and Imbrie \cite{FI}. For other recent work developing their ideas, see
\cite{BC01,BC02}.

\section{Cluster expansion of balanced contours}\label{sec: balanced}

In this section we perform the zeroth step of the multi-scale
analysis for the polymer model~\eqref{eq: polymer model}, and set
up the cluster expansion for a class of contours the weight of
which is sufficiently damped. As a result, an interacting polymer
model is obtained that will be dealt with in the next section.

Let an integer $l_0$ be fixed. It is supposed to be large enough and the precise conditions will be specified throughout the sequel. It plays the role of an
$\eta$-independent 'cut-off scale'.
Given any boundary condition $\eta$ (fixed throughout this
section), we start by defining the set of contours that allow for the cluster expansion.
Obviously, every bulk contour $\Ga$ has the weight $\rho^\eta(\Ga) = \exp(-2\be|\Ga|)$.
For boundary contours, there is no such exponential bound with a strictly positive rate, uniformly in $\eta$. Instead, we segregate an $\eta$-dependent subset of sufficiently damped boundary contours as follows.
\begin{definition}\label{de: balanced contour}
Given $\eta \in \Om$, a boundary contour $\Ga$ is called balanced (or
$\eta$-balanced) whenever
\begin{equation}
  \sum_{x \in \un{\pa^-\Ga}} \eta_x \geq -\bigl( 1 - \frac{1}{l_0} \bigr) |\Ga|
\end{equation}
Otherwise $\Ga$ is called unbalanced.\\
A set $B \subset \pa\La$ is called unbalanced if there exists an unbalanced contour
$\Ga$, $\pa^-\Ga = B$.
\end{definition}
While the case of large boundary contours will be discussed separately in the next section, some basic properties of unbalanced small boundary contours are collected in the following lemma. We define the \emph{height} of any simple boundary contour $\Ga$ as
\begin{equation}
  h(\Ga) = \max_{y^* \in \Ga} d[y^*,\pa\Ga]
\end{equation}
In order to extend this definition to small boundary contours
$\Ga$ such that $\pa\Ga$ contains an (exactly one) corner, we make
the following construction. If
$\pa\Ga$ is a connected subset of the boundary with the endpoints
$[\pm (N+1/2),a]$ and $[b,\pm (N+1/2)]$, then we define
the set $R(\Ga) \subset (\bbZ^2)^*$ as the (unique) rectangle such
that $[\pm (N+1/2),a]$, $[b,\pm (N+1/2)]$, and
$[\pm (N+1/2),\pm (N+1/2)]$ are
three of its corners. Now the height is the maximal distance of a
point in the contour to this rectangle,
\begin{equation}
  h(\Ga) = \max_{y^* \in \Ga} d[y^*,R(\Ga)]
\end{equation}
The situation is illustrated in Figure~6.

\begin{figure}
\begin{center}
\includegraphics[scale=0.6]{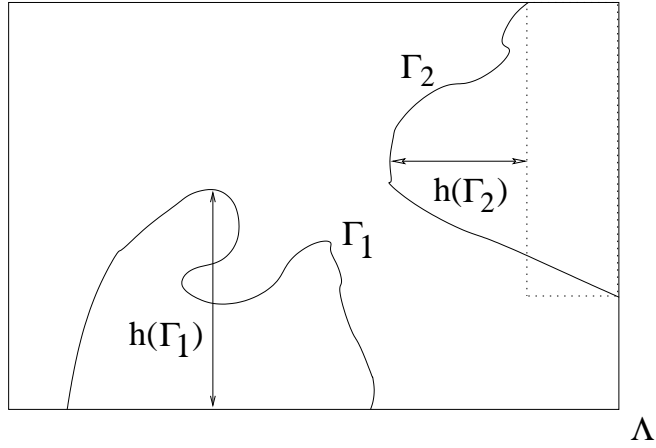}
\caption{Height of small boundary contours.}
\end{center}
\end{figure}

\begin{lemma}\label{lem: geom-balanced}
Let $\Ga$ be an unbalanced small boundary contour. Then,
\begin{enumerate}
\item[i)]
$\sum_{x \in \un{\pa\Ga}} \eta_x \leq -\bigl( 1 - \frac{2}{l_0} \bigr) |\pa\Ga|$.

\item[ii)]
$|\pa\Ga| \geq l_0 h(\Ga)$. In particular, if $\Ga$ is simple then
$|\pa\Ga| \geq l_0$.
\end{enumerate}
\end{lemma}
\begin{proof}
For any unbalanced contour $\Ga$, Definition~\ref{de: balanced
contour} together with the bound $|\Ga| \geq |\pa\Ga|$ valid for any
small boundary contour implies the inequalities
\begin{equation}\label{eq: multi-ineq}
  -|\pa^-\Ga| \leq \sum_{x \in \un{\pa^-\Ga}} \eta_x
  < -\bigl( 1 - \frac{1}{l_0} \bigr) |\Ga|
  \leq -\bigl( 1 - \frac{1}{l_0} \bigr) |\pa\Ga|
\end{equation}
Hence, $|\pa^+\Ga| \leq \frac{1}{l_0}|\pa\Ga|$ and we obtain
\begin{equation}
  \sum_{x \in \un{\pa\Ga}} \eta_x \leq -\bigl( 1 - \frac{1}{l_0} \bigr) |\pa\Ga|
  + |\pa^+\Ga| \leq -\bigl( 1 - \frac{2}{l_0} \bigr) |\pa\Ga|
\end{equation}
proving i).

If $\Ga$ is simple, then we use \eqref{eq: multi-ineq} again together
with the refined relation $|\Ga| \geq |\pa\Ga| + 2h(\Ga)$ to get
\begin{equation}\label{eq: trivial}
  |\pa\Ga| \geq \bigl( 1 - \frac{1}{l_0} \bigr) |\Ga|
  \geq \bigl( 1 - \frac{1}{l_0} \bigr) (|\pa\Ga| + 2h(\Ga))
\end{equation}
which implies $|\pa\Ga| \geq l_0 h(\Ga) \geq l_0$, assuming $l_0
\geq 2$ and using that
$h(\Ga) \geq 1$ for any simple small boundary contour. \\
Since the definition of the height is such that the inequality
$|\Ga| \geq |\pa\Ga| + 2h(\Ga)$ remains true as well for any small boundary contour
$\Ga$ such that $\pa\Ga$ contains a corner, the lemma is proven.
\end{proof}

The union of the set of all bulk contours and of the set of all balanced boundary contours is denoted by $\caK_0^\eta$. We also write $\scD_0^\eta$ for the set of all compatible families of contours from $\caK_0^\eta$, and $\scD_{>0}^\eta$ for the set of all compatible families of contours from $\caK_\La \setminus \caK_0^\eta$.
Later we will show that, for almost every $\eta$, all large boundary contours (i.e.\ those containing at least one interface) are balanced for all but finitely many squares $\La(N)$.

Formally, the partition function
\eqref{eq: polymer model} can be partially computed by
summing over all contours from the set
$\caK_0^\eta$. We start by rewriting partition function~\eqref{eq: polymer model} as
\beq
  \caZ_\La^{\eta} = \exp{(-E^\eta(\emptyset))}
  \sum_{\pa\in\scD_{>0}^\eta}
  \prod_{\Ga \in \partial} \rho^\eta(\Ga)
  \sum_{\pa^0 \in \scD_0^\eta \atop \pa^0 \sim \pa}
  \prod_{\Ga^0 \in \partial^0} \rho^\eta(\Ga^0)
\eeq Here, the first sum runs over all compatible families $\pa$ of
contours not belonging to $\caK_0^\eta$, while the second one is over
all collections of contours from $\caK_0^\eta$, compatible with
$\pa$. Let $\frC_\La^0$ denote the set of all clusters of contours
from $\caK_0^\eta$. Then, the cluster expansion reads, see
Appendix~\ref{ap: cluster models},

\begin{equation}
\caZ_\La^{\eta} = \exp{(-E^\eta(\emptyset))}
  \sum_{\partial \in \scD_{>0}^\eta}
  \prod_{\Ga \in \partial} \rho^\eta(\Ga)
  \exp{\Bigl(\sum_{C \in \frC_0^\eta \atop C \sim \partial}
\phi_0^\eta(C)\Bigr)}
\end{equation}
where the sum runs over all clusters of contours from $\caK_0^\eta$
that are compatible with $\partial$, and we have denoted the weight
of a cluster $C$ by $\phi_0^\eta(C)$.  Note that the cluster
expansion was applied only formally here and it needs to be justified
by providing bounds on the cluster weights. This is done in
Proposition~\ref{le: clusters step zero} below.

Hence, we rewrite the model with the partition function $\caZ_\La^{\eta}$ as an effective model upon the contour ensemble $\caK_\La \setminus \caK_0^\eta$, with a contour interaction mediated by the clusters:
\begin{align}\label{Z1}
  \caZ_\La^\eta=&\caZ_1^\eta\, \exp{\Bigl(-E^\eta(\emptyset)+\sum_{C \in \frC_0^\eta}
  \phi_0^\eta(C)\Bigr)}
\\\intertext{where}
  \label{Z1'}
  \caZ_1^\eta=&\sum_{\partial\in\scD_{>0}^\eta}
  \exp{\Bigl(-\sum_{C \in \frC_0^\eta \atop C \not\sim \partial}
  \phi_0^\eta(C)\Bigr)}
  \prod_{\Ga \in \partial} \rho^\eta(\Ga)
\end{align}
After establishing an exponential upper bound on the number of incompatible contours in the next lemma, a bound on the cluster weights immediately follows by recalling
the basic result on the convergence of the cluster expansions~\cite{KoPr}.
\begin{lemma}\label{lem: entropy}
There exists a constant $c_1 > 0$ (independent of $l_0$) such that the number of all contours $\Ga' \in \caK_\La$, $|\Ga'| = n$, $\Ga' \not\sim \Ga$ is upper-bounded by $|\Ga|\,e^{c_1 n}$, for any $\Ga \in \caK_\La$ and $n = 1,2,\ldots$
\end{lemma}
\begin{proof}
Note that $\Ga$ is not necessarily a connected set. However,
the relation $\Ga' \nsim \Ga$ implies
$(\Ga' \cup \pa\Ga') \cap (\Ga \cup \pa\Ga) \neq \emptyset$, and using that
$\Ga \cup \pa\Ga$ is connected, we get:
\begin{equation*}
\begin{split}
  \#\{\Ga':\,\Ga' &\nsim \Ga,\,|\Ga'| = n\} \leq
  |\Ga \cup \pa\Ga|\,\sup_{x^*}\,\#
  \{\Ga':\, x^* \in \Ga' \cup \pa\Ga',\,|\Ga'| = n\}
\\
  &\leq 3|\Ga|\,\sup_{x^*} \#\{A \subset (\bbZ^2)^* \text{ connected},\,
  x^* \in A,\,|A| \leq 3n \}
\\
  &\leq |\Ga|\,\cdot 4^{6n+1} \leq |\Ga|\,e^{c_1 n}
\end{split}
\end{equation*}
by choosing $c_1$ large enough.
\end{proof}

Assigning to any cluster $C \in \frC_0^\eta$ the \emph{domain}
$\dom(C) = \un{\pa C}$ where $\pa C = \cup_{\Ga \in C} \pa\Ga$ is the boundary of $C$,
and the length $|C| = \sum_{\Ga \in C} |\Ga|$, we have the following
result.
\begin{proposition}\label{le: clusters step zero}
There are constants $\be_4,c_2 > 0$ (independent of $l_0$) such that for any
$\be \geq l_0 \be_4$, one has the upper bound
\begin{equation}\label{eq: clusters step zero}
  \sup_{x^*} \sum_{C \in\frC_0^\eta \atop x^* \in C}
  |\phi_0^\eta|\,\exp{\Bigl[ \bigl(\frac{2\be}{l_0} - c_2 \bigr)|C|
  \Bigr]} \leq 1
\end{equation}
uniformly in $\La$.\\
Moreover, $\phi_0^\eta(C)$ only depends on the restriction of $\eta$ to the set
$\dom(C)$.
\end{proposition}
\begin{proof}
Using Definition~\ref{de: balanced contour} and equation~\eqref{eq: rhoGa1},
we have $\rho^\eta(\Ga) \leq \exp(-\frac{2\be}{l_0}|\Ga|)$ for any balanced contour $\Ga$. In combination with Lemma~\ref{lem: entropy}, we get
\begin{equation}
  \sum_{\Ga \in \caK_0^\eta \atop x^* \in \Ga} |\rho^\eta(\Ga)|\,
  \exp\bigl[ (\frac{2\be}{l_0} - c_2 + 1 )|\Ga\|\bigr] \leq
  \sum_{n=1}^{\infty} \exp[ -(c_2 - c_1 - 1)n ] \leq 1
\end{equation}
provided that $c_2$ is chosen large enough. The proposition now
follows by applying Proposition~\ref{prop: cluster model}, with
$\be_4 = \frac{c_2}{2}$.
\end{proof}

\section{Absence of large boundary contours}\label{sec: large contours}

By the construction, all unbalanced contours are boundary contours, either small or large. In this section we show that unbalanced large boundary contours actually do not exist under a typical realization of the boundary condition. This observation will allow us to restrict our multi-scale analysis entirely to the class of small boundary contours.
\begin{lemma}\label{lem: geom-large}
There is a constant $c_3 > 0$ such that for any $N \in \bbN$ and any unbalanced large boundary contour $\Ga \in \caK_{\La(N)}$, the inequality
\begin{equation}
  \sum_{x \in \un{\pa\Ga}} \eta_x \leq -c_3 N
\end{equation}
holds true.
\end{lemma}
\begin{proof}
Using the geometrical inequality
$|\Ga| \geq 2N + |\pa^+\Ga|$ and Definition~\eqref{de: balanced contour}, we have
\begin{equation}
\begin{split}
  \sum_{x \in \un{\pa\Ga}} \eta_x &\leq -\bigl( 1-\frac{1}{l_0} \bigr) |\Ga|
  + \sum_{x \in \un{\pa^+\Ga}} \eta_x
\\
  &\leq -\bigl( 1-\frac{1}{l_0} \bigr) (2N + |\pa^+\Ga|) + |\pa^+\Ga|
\\
  &\leq -2N \bigl( 1-\frac{3}{l_0} \bigr)
\end{split}
\end{equation}
where in the last inequality we used that
$|\pa^+\Ga| \leq |\pa\Ga| \leq 4N$.
\end{proof}
\begin{proposition}
There is a constant $c_4 > 0$ such that for any $N \in \bbN$,
\begin{equation}
  \bsP\{\exists\Ga \in \caK_{\La(N)}\text{ large unbalanced}\} \leq
  \exp (-c_4 N)
\end{equation}
\end{proposition}
\begin{proof}
If $B \subset \pa\La(N)$ is a connected set containing exactly two corners, then, using
Lemma~\ref{lem: geom-large},
\begin{equation}
\begin{split}
  \bsP\{\exists \Ga \in \caK_{\La(N)} \text{ large unbalanced}:\, \pa\Ga = B\}
  &\leq \bsP \bigl\{ \sum_{x \in \un{B}} \eta_x \leq -c_3 N \bigr\}
\\
  \leq\bsP \bigl\{ \sum_{x \in \un{B}} \eta_x \leq -\frac{c_3}{2} |B| \bigr\}
  &\leq \exp \bigl( -\frac{c_3^2}{8} |B| \bigr)
\end{split}
\end{equation}
Hence,
\begin{equation}
\begin{split}
  \bsP\{\exists\Ga &\in \caK_{\La(N)}\text{ large unbalanced}\}
\\
  &\leq \sum_{B \subset \pa\La}
  \bsP\{\exists \Ga \in \caK_{\La(N)} \text{ large unbalanced}:\, \pa\Ga = B\}
\\
  &\leq \sum_{l \geq 2N} 8N\,\exp \bigl( -\frac{c_3^2}{8} l \bigr)
  \leq \frac{128N}{c_3^2} \exp \bigl( -\frac{c_3^2 N}{4} \bigr)
  \leq \exp (-c_4 N)
\end{split}
\end{equation}
by choosing $c_4$ large enough.
\end{proof}
\begin{corollary}\label{cor: no large unbalanced}
There exists a set $\Om^* \subset \Om$, $\bsP\{\Om^*\} = 1$ and a function
$N^*: \Om^* \mapsto \bbN$ such that for any b.c.\
$\eta \in \Om^*$ and any volume $\La = \La(N)$, $N \geq N^*(\eta)$, all large boundary contours are balanced.
\end{corollary}
\begin{proof}
Since
\begin{equation}
  \sum_N \bsP\{\exists\Ga \in \caK_{\La(N)}\text{ large unbalanced}\}
  < \infty
\end{equation}
the Borel-Cantelli lemma implies
\begin{equation}
  \bsP\{\forall N_0 \in \bbN:\, \exists N \geq N_0:\
  \exists\Ga \in \caK_{\La(N)}\text{ large unbalanced}\}
  = 0
\end{equation}
proving the statement.
\end{proof}

We are now ready to prove the almost sure absence of interfaces in the large-volume limit.
\begin{proof}[Proof of Theorem~\ref{thm: interfaces}]
Let $\eta \in \Om^* \cap (-\Om^*)$ and $N \geq N^*(\eta)$. Then, any large boundary contour $\Ga$ is both $\eta$- and $(-\eta)$-balanced and, using the Peierls inequality and \eqref{eq: representation-Gibbs}, the Gibbs probability of any collection of (possibly large boundary) contours
$\Ga_1,\ldots,\Ga_m$, $m = 1,2,\ldots$ has the upper bound
\begin{equation}
\begin{split}
  \mu_{\La(N)}^\eta(&\Ga_1,\ldots,\Ga_m)
  \leq \max_{a \in \{-1,1\}} \nu_{\La(N)}^{a\eta}(\Ga_1,\ldots,\Ga_m)
\\
  &\leq \max_{a \in \{-1,1\}} \prod_{i=1}^m \rho^{a\eta}(\Ga_i)
  \leq \exp\bigl( -\frac{2\be}{l_0} \sum_{i=1}^{m}|\Ga_i| \bigr)
\end{split}
\end{equation}
Hence, using Lemma~\ref{lem: entropy} and the bound $|\Ga| \geq 2N$ for any large boundary contour $\Ga$, we get
\begin{equation}
\begin{split}
  \mu_{\La(N)}^\eta(&\exists \text{ a large boundary contour})
  \leq \sum_{m=1}^\infty \frac{1}{m!} \sum_{\Ga_1,\ldots,\Ga_m \text{ large}\atop
  \forall i: \Ga_i \cap \pa\La \neq \emptyset}
  \mu_{\La(N)}^\eta(\Ga_1,\ldots,\Ga_m)
\\
  &\leq \sum_{m=1}^\infty \frac{1}{m!} \sum_{x_1,\ldots,x_m \in \pa\La}
  \sum_{\Ga_1 \ni x_1,\ldots,\Ga_m \ni x_m}
  \exp\bigl( -\frac{2\be}{l_0} \sum_{i=1}^{m}|\Ga_i| \bigr)
\\
  &\leq \exp\bigl( -\frac{2\be}{l_0} N \bigr) \sum_{m=1}^\infty \frac{1}{m!}
  \bigl( 4N\,\sum_{\Ga \ni x \atop |\Ga| \geq 2N}
  \,e^{-\frac{\be}{l_0}|\Ga|} \bigr)^m
\\
  &\leq \exp \bigl[ - \bigl( \frac{2\be}{l_0} -
  8 e^{-2(\frac{\be}{l_0} - c_1)N} \bigr)N \bigr] \longrightarrow 0
\end{split}
\end{equation}
provided that $\be$ is large enough. Since $\bsP\{\Om^* \cap (-\Om^*)\} = 1$, the theorem is proven.
\end{proof}

As a consequence, all interfaces get $\bsP$-a.s.\ and for all but finitely many volumes uniformly exponentially damped weights. Hence, their Gibbs probabilities become exponentially small as functions of the size of the system and, therefore, no interfacial infinite-volume Gibbs measure occurs as a limit point, with $\bsP$-probability 1. While such a result is not sensational in $d = 2$ (in this case, no translationally non-invariant Gibbs measure exists by \cite{Aiz,Hi}), similar arguments  are expected to apply in higher dimensions.

In the next sections, a perturbation technique is developed that allows us to address the question whether non-trivial mixtures of $\mu^+$ and $\mu^-$ can occur as limit measures.

\section{Classification of unbalanced contours}\label{sec: classification}

We now consider the interacting contour model introduced by the
partition function \eqref{Z1'}, defined on the set of
unbalanced contours $\caK_\La \setminus \caK_0^\eta$.
As a consequence of Corollary~\ref{cor: no large unbalanced}, we can restrict our analysis to the set $\Om^*$ of boundary conditions under which the set
$\caK_\La \setminus \caK_0^\eta$ of unbalanced contours contains only small boundary contours, both simple and corner ones.

Our multi-scale analysis consists of a sequential expansion of groups of unbalanced
contours that are far enough from each other. The
groups are supposed to be typically sufficiently rarely
distributed, so that the partition function \eqref{Z1} can be
expanded around the product over the partition functions computed
within these groups only. Under the condition that the density of
the groups decays fast enough with their space extension, one can
arrive at an expansion that essentially shares the locality
features of the usual cluster expansion, at least for
$\bsP$-typical boundary conditions $\eta$.
To make this strategy work, we define a suitable
decomposition of the set $\caK_\La \setminus \caK_0^\eta$ into
disjoint groups associated with a hierarchy of length scales.
Also, the unbalanced contours close enough to any of the four corners
will be dealt with differently and expanded in the end.
\begin{definition}\label{assumptions scales}
Assuming $l_0$ to be fixed, we define the two sequences
$(l_n)_{n=1,2,\ldots}$ and $(L_n)_{n=1,2,\ldots}$ by the following recurrence relations:
\begin{equation}
\begin{split}
  L_n &= \frac{l_{n-1}}{5^n}\,, \quad\quad
  l_n = \exp\bigl(\frac{L_n}{2^n}\bigr) \qquad\qquad n = 1,2,\ldots
\end{split}
\end{equation}
\end{definition}
For any $n = 1,2,\ldots$, any pair of contours $\Ga,\Ga'$ is called $L_n$-connected, if
$d[\Ga,\Ga'] \leq L_n$.
Furthermore, fixing a positive constant $\ep > 0$, we introduce the $N$-dependent length scale
\begin{equation}
  l_\infty = (\log N)^{1 + \ep}
\end{equation}
Introducing the boundary $\pa\De$ for any set of contours
$\De \subset \caK_\La$ by $\pa\De = \cup_{\Ga \in \De} \pa\Ga$,
we consider the $\eta$-dependent decomposition of the set of contours
$\caK_\La \setminus \caK_0^\eta$ defined by induction as follows.
\begin{definition}\label{def: decomposition}
\begin{enumerate}
\item[1)]
A maximal $L_1$-connected subset $\De \subset \caK_\La \setminus \caK_0^\eta$ is called a \emph{$1$-aggregate} whenever i) $|\pa\De|\con \leq l_1$, ii) there is no corner
$x^*_{C,i}$ such that \linebreak $\max_{y^* \in \pa\De} d[y^*,x^*_{C,i}] \leq l_\infty$.
We use the notation $(\caK_{1,\al}^\eta)$ for the collection of all $1$-aggregates, and write $\caK_1^\eta = \cup_\al \caK_{1,\al}^\eta$.
\item[\vdots]
\item[n)]
Assume the sets $(\caK_{j,\al}^\eta)_{j = 1,\ldots,n-1}$ have been defined.
Then, the $n$-aggregates are defined as maximal $L_n$-connected subsets
$\De \subset \caK_\La \setminus \cup_{j<n} \caK_j^\eta$ satisfying
i) $|\pa\De|\con \leq l_n$, ii) there is no corner
$x^*_{C,i}$ such that $\max_{y^* \in \pa\De} d[y^*,x^*_{C,i}] \leq l_\infty$.
The set of all $n$-aggregates is denoted by
$(\caK_{n,\al}^\eta)$, and
$\caK_n^\eta = \cup_\al \caK_{n,\al}^\eta$.
\end{enumerate}
To each $n$-aggregate $\caK_{n,\al}^\eta$ we assign the domain
\begin{equation}
  \dom(\caK_{n,\al}^{\eta}) := \un{\{x^*\in\pa\La;\,d[x^*,\pa\caK_{n,\al}^\eta]
  \leq L_n\}}
\end{equation}
\end{definition}
Obviously, the set
$\caK_\infty^\eta := \caK_\La \setminus (\caK_1^\eta \cup \caK_2^\eta \cup \ldots)$
need not be empty, and since all large boundary contours are
balanced, for every contour $\Ga \in \caK_\infty^\eta$ there is
exactly one corner $x^*_{C,i}$ such that
$\max_{y^* \in \pa\Ga} d[y^*,x^*_{C,i}] \leq l_\infty$.
Hence, there is a natural decomposition of the set $\caK_\infty^\eta$
into at most four \emph{corner aggregates},
$\caK_\infty^\eta = \cup_i \caK_{\infty,i}^\eta$, each of them consisting of contours
within the logarithmic neighborhood of one of the corners. In
general, any corner aggregate contains both simple and corner
boundary contours. Later we will show that with $\bsP$-probability 1,
every unbalanced corner boundary contour belongs to a corner
aggregate. In other words, every $n$-aggregate, $n = 1,2,\ldots$
contains only simple boundary contours.
\begin{remark}
By Definition~\ref{def: decomposition},
any $n$-aggregate has a distance at least $L_n$ from all
$m$-aggregates, $m \geq n$. In this way, in the $n$-th step of our expansion,
after having removed all lower-order aggregates, we will be able to
use the `essential independence' of all $n$-aggregates. Namely, on
the assumption that $L_n$ is big enough, depending on the aggregate
size $l_n$, both the interaction among the $n$-aggregates and the
interaction between $n$-aggregates and $m$-aggregates, $m\geq n$ will
be controlled by a cluster expansion.
\end{remark}

Our first observation is a local property of the above construction,
which will be crucial to keep the dependence of expansion terms to be
defined later depending only on a sufficiently small set of boundary
spins.
\begin{lemma}
Let a set of small boundary contours $\De$ be fixed and assume that
$\eta,\eta' \in \Om$ are such that $\eta_{\dom(\De)} = \eta'_{\dom(\De)}$.
Then, $\De$ is an $n$-aggregate w.r.t. the boundary condition $\eta$
if and only if it is an $n$-aggregate w.r.t.\ $\eta'$.
\end{lemma}

The super-exponential growth of the scales $l_n$ will imply an
exponential decay of the probability for an $n$-aggregate to occur.
An upper bound on this probability is stated in the following
proposition, the proof of which is given in Section~\ref{sec:
prob-bound-proof}.
\begin{proposition}\label{prop: prob-bound}
There is a constant $c_5 > 0$ (independent of $l_0$) such that
for any $n = 1,2,\ldots$ and any connected set $B \subset \pa\La$,
\begin{equation}
  \bsP\{\exists \caK_{n,\al}^\eta:\,\cn(\pa\caK_{n,\al}^\eta) = B \}
  \leq e^{-c_5 |B|}
\end{equation}
uniformly in $\La$.
\end{proposition}
Note that, given a connected set $B \subset \pa\La$, there is at most one aggregate
$\caK_{n,\al}^\eta$, $n = 1,2,\ldots$ such that $\cn(\pa\caK_{n,\al}^\eta) = B$.
\begin{corollary}
There exists $\Om^{**} \subset \Om^*$, $\bsP\{\Om^{**}\} = 1$
and $N^{**}: \om^{**} \mapsto \bbN$, $N^{**}(\om) \geq N^*(\om)$ such that
for any $\om \in \Om^{**}$ and any $\La = \La(N)$, $N \geq N^{**}(\om)$
every aggregate $\caK_{n,\al}^\eta$, $n = 1,2,\ldots$
satisfies the inequality $|\pa\caK_{n,\al}^\eta|\con \leq l_\infty$.
In particular:
\begin{enumerate}
\item[i)]
The set $\cn(\caK_{n,\al}^\eta)$ is a boundary interval and there is at most one corner
$x^*_{C,i}$ such that $d[x^*_{C,i},\pa\caK_{n,\al}^\eta] \leq l_\infty$.
\item[ii)]
All contours $\Ga \in \caK_{n,\al}^\eta$ are simple boundary contours.
\end{enumerate}
\end{corollary}
\begin{proof}
Using Proposition~\ref{prop: prob-bound}, the probability for any aggregate to occur can be estimated as
\begin{equation}  \label{eq: prob-bound1}
\begin{split}
  \bsP\{\exists &\caK_{n,\al}^\eta,\,n=1,2,\ldots:\,|\pa\caK_{n,\al}^\eta|\con >
  l_\infty \} \leq
  \sum_{B \subset \pa\La \text{ conn.} \atop |B| > (\log N)^{1+\ep}}
  \bsP\{\exists \caK_{n,\al}^\eta:\, \cn(\caK_{n,\al}^\eta) = B\}
\\
  &\leq |\pa\La| \sum_{l > (\log N)^{1+\ep}} e^{-c_5 l} \leq \frac{16}{c_5}
  N^{1-c_5(\log N)^\ep} = o\bigl( N^{-\de} \bigr)
\end{split}
\end{equation}
for any (arbitrarily large) $\de > 0$. Hence,
\begin{equation}\label{eq: prob-to-prove}
  \sum_{N=1}^{\infty} \bsP\{ \exists \caK_{n,\al}^\eta,\, n=1,2,\ldots:\,
  |\pa\caK_{n,\al}^\eta|\con > l_\infty \} < \infty
\end{equation}
and the statement follows by a Borel-Cantelli argument.
\end{proof}

For convenience, let us summarize the results of the last three sections by reviewing
all types of contours again together with their balancedness properties. For any
$\eta \in \Om^{**}$ and $\La = \La(N)$, $N \geq N^{**}(\om)$, any configuration of contours
$\pa \in \scD_\La$ possibly contains
\begin{enumerate}
\item[i)]
\emph{Bulk contours} (trivially balanced).
\item[ii)]
\emph{Large boundary contours} that are balanced.
\item[iii)]
\emph{Corner boundary contours} that are either balanced or elements of corner aggregates.
\item[iv)]
\emph{Simple boundary contours} which are balanced or elements of either
$n$-aggre\-gates, $n = 1,2,\ldots$, or of corner aggregates.
\end{enumerate}

\section{Sequential expansion of unbalanced contours}\label{sec: sequential}

The next step in our strategy is to proceed by induction in the order
of aggregates, rewriting at each step the interacting polymer
model~\eqref{Z1'} as an effective model over the contour ensembles
$\caK_\La \setminus (\caK_0^\eta \cup \caK_1^\eta)$,
$\caK_\La \setminus (\caK_0^\eta \cup \caK_1^\eta \cup \caK_2^\eta)$, etc.
At the $n$-th step, a compatible set of contours inside all corner
and all normal $m$-aggregates, $m > n$, is fixed, and we perform the
summation over contours in all normal $n$-aggregates. This is a
constrained partition function which is approximately a product over
the normal $n$-aggregates.  By the construction, the latter are
sufficiently isolated on the scale $L_n$, which will allow for the
control of the remaining interaction by means of a cluster expansion.
At the end, we arrive at an effective model over the contour ensemble
$\caK_\infty^\eta$, which is the union of (at most four) corner
aggregates. In large volumes, the corner aggregates become
essentially independent, the error being exponentially small in the
size of the volume. The reason we distinguish between the
$n$-aggregates and the corner aggregates is that the partition
function within the former allows for a much better control, which
will be essential in our analysis of the characteristic function of
the random free energy difference
$\log \caZ_\La^\eta - \log \caZ_\La^{-\eta}$ in Section~\ref{sec: prob-analysis}.
Note that the lack of detailed control around the corners is to be
expected as there may more easily occur some low-energy (unbalanced)
boundary contours, but at most of logarithmic size in $N$.

The $n$-th step of the expansion, $n \geq 1$, starts from the
partition function, \begin{equation}\label{eq: pf-induction}
  \caZ_n^\eta = \sum_{\pa \in \scD_{>n-1}^\eta} \exp \Bigl(
  -\sum_{C \in \frC_{n-1}^\eta \atop C \nsim \pa} \phi_{n-1}^\eta(C) \Bigr)
  \prod_{\Ga \in \pa} \rho^\eta(\Ga)
\end{equation}
which in the case $n=1$ coincides with~\eqref{Z1'}. Here,
$\scD_{>n-1}^\eta$ is the set of all compatible families of contours
from $\caK_{>n-1}^\eta := \caK_\La^\eta \setminus (\caK_0^\eta \cup
\caK_1^\eta \cup \ldots \cup \caK_{n-1}^\eta)$, i.e.\ with all normal
$m$-aggregates, $m \leq n-1$, being removed. Furthermore, we use the
notation $\frC_{n-1}^\eta$ for the set of all $(n-1)$-clusters. Here,
the $0$-clusters have been introduced in Section~\ref{sec: balanced},
and the clusters of higher order will be defined inductively in the
sequel.

In order to analyze partition function~\eqref{eq: pf-induction}, we
follow the ideas of Fr\"ohlich and Imbrie \cite{FI}, however, we
choose to present them in a slightly different way. Observing that,
by construction, the family of aggregates compose a `sparse set', one
is tempted to approximate the partition function by a product over
the aggregates and to control the error by means of a cluster
expansion. However, to make this strategy work, we need to
`renormalize' suitably the contour weights. Namely, only the clusters
that intersect at least two distinct aggregates generate an
interaction between them, and are sufficiently damped by using the
sparsity of the set of aggregates. On the other hand, the
(sufficiently short) clusters intersecting a single aggregate cannot
be expanded, and they modify the weights of contour configurations
within the aggregate. An important feature of this procedure is that
the weight of these contour configurations is kept positive. In some
sense, it is this very renormalization of the weights within each
aggregate that can hardly be done via a single expansion and requires
an inductive approach. In what follows, we present this strategy in
detail, via a number of steps.

\subsection{Renormalization of contour weights}

For any compatible set of contours $\pa \subset \caK_n^\eta$, define the renormalized
weight
\begin{equation}
  \hat\rho^\eta(\pa) =
  \exp \Bigl( -\sum_{C \in \frC_{n-1}^\eta \atop C \nsim \pa;\, |C| < L_n}
  \phi_{n-1}^\eta(C)  \Bigr) \prod_{\Ga \in \pa} \rho^\eta(\Ga)
\end{equation}
Note that the above sum only includes the clusters of length smaller
than $L_n$. By construction, any such cluster is incompatible with at
most one $n$-aggregate. Hence, the renormalized weight
$\hat\rho^\eta(\Ga)$ factorizes over the $n$-aggregates and we have
$\hat\rho^\eta(\pa) = \prod_\al \hat\rho^\eta(\pa^\al)$ where
$\pa^\al = \pa \cap \caK_{n,\al}^\eta$.
Therefore, formula~\eqref{eq: pf-induction} gets the form
\begin{equation}
\begin{split}
  \caZ_n^\eta &= \sum_{\pa \in \scD_{>n}^\eta}
  \prod_{\Ga \in \pa} \rho^\eta(\Ga)
  \sum_{\pa^n \in \scD_n^\eta} \hat\rho^\eta(\pa^n)
  \exp \Bigl(
  -\sum_{C \in \frC_{n-1}^\eta \atop (C \nsim \pa) \vee (C \nsim \pa^n;\, |C| \geq L_n)}
  \phi_{n-1}^\eta(C) \Bigr)
\\
  &= \sum_{\pa \in \scD_{>n}}
  \prod_{\Ga \in \pa} \rho^\eta(\Ga)
  \exp \Bigl(  -\sum_{C \in \frC_{n-1}^\eta \atop C \nsim \pa} \phi_{n-1}^\eta(C) \Bigr)
\\
  &\phantom{=  } \times \sum_{\pa^n \in \scD_n^\eta} \hat\rho^\eta(\pa^n)
  \exp \Bigl(  -\sum_{C \in \frC_{n-1}^\eta;\,|C| \geq L_n \atop
  C \sim \pa;\, C \nsim \pa^n} \phi_{n-1}^\eta(C) \Bigr)
\end{split}
\end{equation}

Defining the renormalized partition function $\hat\caZ_{n,\al}^\eta$ of the contour ensemble $\caK_\La^{n,\al}$ as
\begin{equation}
  \hat\caZ_{n,\al}^\eta = \sum_{\pa^n \in \scD_{n,\al}^\eta} \hat\rho^\eta(\pa^n)
\end{equation}
and using the shorthand
\begin{equation}
  \tilde\phi_{n-1}^\eta(C,\pa^n) = \phi_{n-1}^\eta(C) \funit_{\{C \nsim \pa^n;\,|C| \geq L_n\}}
\end{equation}
we obtain
\begin{equation}\label{eq: induction1}
\begin{split}
  \caZ_n^\eta = &\prod_\al \hat\caZ_{n,\al}^\eta\,\sum_{\pa \in \scD_{>n}^\eta}
  \prod_{\Ga \in \pa} \rho^\eta(\Ga)
  \exp \Bigl(  -\sum_{C \in \frC_{n-1}^\eta \atop C \nsim \pa} \phi_{n-1}^\eta(C) \Bigr)
\\
  &\times \sum_{\pa^n \in \scD_n^\eta}
  \prod_\al \frac{\hat\rho^\eta(\pa^{n,\al})}{\hat\caZ_{n,\al}^\eta}
  \exp \Bigl(  -\sum_{C \in \frC_{n-1}^\eta \atop C \sim \pa}
  \tilde\phi_{n-1}^\eta(C,\pa^n) \Bigr)
\end{split}
\end{equation}
where $\pa^{n,\al} = \pa^n \cap \caK_{n,\al}^\eta$ is the restriction of $\pa^n$ to the
$n$-aggregate $\caK_{n,\al}^\eta$. In the last expression, the second sum contains the interaction between $n$-aggregates, to make a correction to the product over the renormalized partition functions $\hat\caZ_{n,\al}^\eta$.

\subsection{Cluster expansion of the interaction between $n$-aggregates}

Now we employ a trick familiar from the theory of high-temperature (Mayer) expansions,
and assign to any family $\caC \subset \frC_{n-1}^\eta$ of
$(n-1)$-clusters the weight
\begin{equation}
  w_n^\eta(\caC) = \frac{1}{\prod_\al \hat\caZ_\La^{n,\al}}
  \sum_{\pa^n \in \scD_n^\eta} \hat\rho^\eta(\pa^n)
  \prod_{C \in \caC} \bigl( e^{-\tilde\phi_{n-1}^\eta(C,\pa^n)} - 1 \bigr)
\end{equation}
See Figure~7 for an example of a family of 1-clusters that
generically gets a nontrivial weight according to this construction.

\begin{definition}
Any pair of $(n-1)$-clusters $C_1,C_2 \in \frC_{n-1}^\eta$ is called
\emph{$n$-incompatible}, $C_1\stackrel{n}{\not\lra} C_2$, whenever
there exists an $n$-aggregate $\caK_{n,\al}^\eta$ such that
$C_1 \nsim \caK_{n,\al}^\eta$ and $C_2 \nsim \caK_{n,\al}^\eta$. \\
In general, the sets $\caC_1,\caC_2 \subset \frC_{n-1}^\eta$ are
\emph{$n$-incompatible} if there are $C_1 \in \caC_1$, $C_2 \in \caC_2$,
$C_1 \stackrel{n}{\not\lra} C_2$.
\end{definition}
One easily checks the following properties of the weight $w_n^\eta(\caC)$.
\begin{lemma}\label{lem: n-cluster}
For any set of $(n-1)$-clusters $\caC \in \frC_{n-1}^\eta$,
\begin{enumerate}
\item[i)] $\sup_\eta |w_n^\eta(\caC)| \leq \prod_{C \in \caC}
(e^{|\phi_{n-1}^\eta(C)|} - 1)$. \item[ii)] If $\caC = \caC_1 \cup
\caC_2$ such that $\caC_1 \stackrel{n}{\lra} \caC_2$, then
$w_n^\eta(\caC) = w_n^\eta(\caC_1)\,w_n^\eta(\caC_2)$. \item[iii)]
The weight $w_n^\eta(\caC)$ depends only on the restriction of
$\eta$ to the set $(\cup_{C \in \caC} \dom(C)) \cup (\cup'_\al
\dom(\caK_{n,\al}^\eta)$ where the second union is over all
$n$-aggregates $\caK_{n,\al}^\eta$ such that $\caC \nsim
\caK_{n,\al}^\eta$.
\end{enumerate}
\end{lemma}

\begin{figure}
\begin{center}
\includegraphics[scale=0.6]{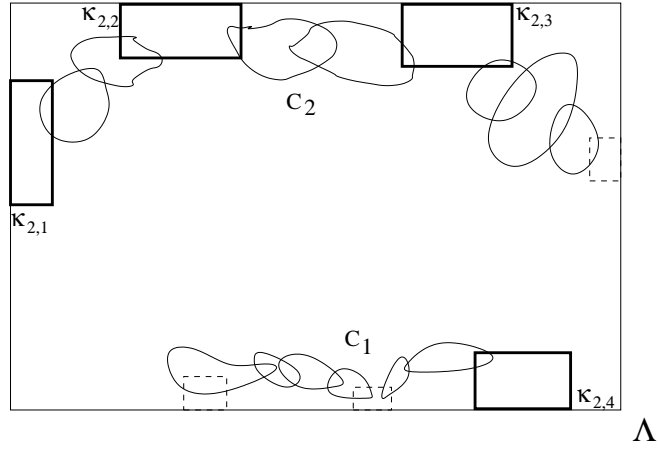}
\caption{A pair of 2-compatible families of 1-clusters $\caC_1, \caC_2 \subset \frC_1^\eta$
intersecting 2-aggregates $K_{2,\al}^\eta$, $\al = 1,2,3,4$. The
dashed rectangles illustrate 1-aggregates which have become parts of
the 1-clusters. By construction,
$w_2^\eta(\caC_1 \cup \caC_2) = w_2^\eta(\caC_1)\,w_2^\eta(\caC_2)$.}
\end{center}
\end{figure}

In the second sum in~\eqref{eq: induction1} we recognize the
partition function of a polymer model with the polymers being defined
as the $n$-connected subsets of
$\frC_{n-1}^\eta$, which are incompatible if and only if they are $n$-incompatible.
Treating this polymer model by the cluster expansion, and using the symbols
$\frD_n^\eta$ for the set of all clusters in this polymer model and
$\psi_n^\eta(D)$ for the weight of a cluster $D \in \frD_n^\eta$ , we get
\begin{equation}\label{eq: induction2}
\begin{split}
  \caZ_n^\eta = &\prod_\al \hat\caZ_{n,\al}^\eta\,\sum_{\pa \in \scD_{>n}^\eta}
  \prod_{\Ga \in \pa} \rho^\eta(\Ga)
  \exp \Bigl(  -\sum_{C \in \frC_{n-1}^\eta \atop C \nsim \pa} \phi_{n-1}^\eta(C) \Bigr)
  \sum_{\caC \subset \frC_{n-1}^\eta \atop \caC \sim \pa} w_n^\eta(\caC)
\\
  &= \exp \Bigl( \sum_{D \in \frD_n^\eta} \psi_n^\eta(D) \Bigr)
  \prod_\al \hat\caZ_{n,\al}^\eta\,\sum_{\pa \in \scD_{>n}^\eta}
  \prod_{\Ga \in \pa} \rho^\eta(\Ga)
\\
  &\phantom{xxx}\times \exp
  \Bigl( -\sum_{C \in \frC_{n-1}^\eta \atop C \nsim \pa} \phi_{n-1}^\eta(C)
  -\sum_{D \in \frD_n^\eta \atop D \nsim \pa} \psi_n^\eta(D) \Bigr)
\end{split}
\end{equation}
Defining the set of all $n$-clusters
$\frC_n^\eta = \frC_{n-1}^\eta \cup \frD_n^\eta$ and the weight of any $n$-cluster
$C \in \frC_n^\eta$ as
\begin{equation}\label{eq: definition n-cluster weights}
  \phi_n^\eta(C) =
  \begin{cases}
    \phi_{n-1}^\eta(C)  &  \text{if } C \in \frC_{n-1}^\eta
  \\
    \psi_n^\eta(C)  &  \text{if } C \in \frD_n^\eta
  \end{cases}
\end{equation}
we finish the inductive step by obtaining the final expression
\begin{equation}
  \caZ_n^\eta = \caZ_{n+1}^\eta
  \prod_\al \hat\caZ_{n,\al}^\eta\,
  \exp \Bigl( \sum_{D \in \frD_n^\eta} \psi_n^\eta(D) \Bigr)
\end{equation}
with the partition function of a new interacting polymer model
\begin{equation}
  \caZ_{n+1}^\eta = \sum_{\pa \in \scD_{>n}^\eta}
  \exp \Bigl( -\sum_{C \in \frC_{n}^\eta \atop C \nsim \pa} \phi_{n}^\eta(C) \Bigr)
  \prod_{\Ga \in \pa} \rho^\eta(\Ga)
\end{equation}

We need to extend the notion of domain from the set of $(n-1)$-clusters $\frC_{n-1}^\eta$ to the set of $n$-clusters $\frC_n^\eta$. Realizing that any $n$-cluster
$D \in \frD_n^\eta$ is a collection $(\caC_i)$ of $L_n$-connected families of
$(n-1)$-clusters, $\caC_i = (C_i^s)$, we first introduce the domain of any such family
$\caC_i$ as $\dom(\caC_i) = \cup_s \dom(C_i^s)$. Next, we define
\begin{equation}
  \dom(D) := \bigcup_i \dom(\caC_i) \cup \
  \bigcup_{\al:\,\caK_{n,\al}^\eta \nsim D} \dom(\caK_{n,\al}^\eta)
\end{equation}
Furthermore, the length $|D|$ of the cluster is defined as
\begin{equation}
  |D| := \sum_i |\caC_i| = \sum_i \sum_s |C_i^s|
\end{equation}
Note that this is possibly much smaller than the diameter of the cluster, since the sizes of the $n$-aggregates in the domain of $D$ are not counted in the length of the cluster. The reason for this definition is that the cluster weights are not expected to be exponentially damped with the cluster diameter. Note, however, that the \emph{probability} of a cluster to occur is exponentially damped with the size of the $n$-aggregates in its domain.

 In the next proposition, we provide uniform bounds on the $n$-cluster weights. For the proof, see Section~\ref{sec: proofs}.
\begin{proposition}\label{Mayer}
There is $\be_5 > 0$ such that for any $\be \geq l_0 \be_5$,
$\eta \in \Om^{**}$, $\La = \La(N)$, $N \geq N^{**}(\eta)$, the inequalities
\begin{align}\label{eq: multi-scale-bound}
  \sup_{x^*} \sum_{D \in \frD_{n}^\eta \atop x^* \in D}
  \exp \bigl( \frac{\be}{l_0}|D| \bigr)\, |\psi_n^\eta(D)|
  &\leq 2^{-n} \qquad n = 1,2,\ldots
\\\intertext{and}\label{eq: multi-scale-bound-phi}
  \sup_n \sup_{x^*} \sum_{C \in \frC_{n}^\eta \atop x^* \in C}
  \exp \bigl( \frac{\be}{l_0}|C| \bigr)\, |\phi_n^\eta(C)|
  &\leq 1
\end{align}
hold true. \\
Moreover, if $C \in \frC_n^\eta$ and $\eta'|_{\dom(C)} = \eta|_{\dom(C)}$, then
also $C \in \frC_n^{\eta'}$ and $\phi_n^{\eta'}(C) = \phi_n^{\eta}(C)$.
Similarly, $D \in \frD_n^\eta$ and $\eta'|_{\dom(D)} = \eta|_{\dom(D)}$ implies
both $D \in \frD_n^{\eta'}$ and $\psi_n^{\eta'}(D) = \psi_n^{\eta}(D)$.
\end{proposition}

\subsection{Expansion of corner aggregates}\label{sec: corners}

For any finite square $\La = \La(N)$ and $\eta \in \Om$, all
aggregates from the set $\cup_n \caK_n^\eta$ are expanded in a finite
number of steps. Afterwards, all corner aggregates are treated by a
similar procedure. Throughout this section, we use the notation $n_0$
for the highest order in the collection of all normal aggregates. The
expansion goes similarly as in the case of normal aggregates, so we
only sketch it.

The renormalized weight of any compatible family of contours
$\pa \subset \caK_\infty^\eta$
is defined by the formula
\begin{equation}
  \hat\rho^\eta(\pa) =
  \exp \Bigl(
  -\sum_{C \in \frC_{n_0}^\eta \atop C \nsim \pa;\, |C| < 2l_\infty}
  \phi_{n_0 - 1}^\eta(C)  \Bigr) \prod_{\Ga \in \pa} \rho^\eta(\Ga)
\end{equation}
which factorizes over the corners,
$\hat\rho^\eta(\pa) = \prod_i \hat\rho^\eta(\pa \cap \caK_{\infty,i}^\eta)$, assuming
$\La(N)$ to be large enough. Clusters $C_1,C_2 \subset \frC_{n_0}^\eta$ are called $\infty$-incompatible whenever there is a corner aggregate $\caK_{\infty,i}^\eta$ such that $C_1 \nsim \caK_{\infty,i}$ and $C_2 \nsim \caK_{\infty,i}$. Defining the weight $w^\eta(C)$ as
\begin{equation}
  w^\eta(\caC) = \frac{1}{\prod_i \hat\caZ_{\infty,i}^\eta}
  \sum_{\pa \in \scD_\infty^\eta} \hat\rho^\eta(\pa)
  \prod_{C \in \caC_{n_0}^\eta}
  \bigl( e^{-\tilde\phi_{n_0}^\eta(C,\pa)} - 1 \bigr)
\end{equation}
where
\begin{align}
  \hat\caZ_{\infty,i}^\eta &= \sum_{\pa \in \scD_{\infty,i}^\eta} \hat\rho^\eta(\pa)
\\    \intertext{and}
  \tilde\phi_\infty^\eta(C,\pa) &=
  \phi_{\infty}^\eta(C) \funit_{\{C \nsim \pa;\,|C| \geq 2l_\infty\}}
\end{align}
an obvious variant of Lemma~\ref{lem: n-cluster} holds true and
$w^\eta(\caC)$ factorizes into a product over maximal connected components of $\caC$ w.r.t.\ $\infty$-incompatibility. Treating these as polymers in a new polymer model with  $\infty$-incompatibility used as the incompatibility relation, and using the notation
$\frD_\infty^\eta$ for the set of all clusters in this polymer model and $\psi_\infty^\eta(D)$ for the cluster weights, we obtain as the final step of the sequential expansion,
\begin{equation}
  \caZ_{n_0+1}^\eta = \exp \Bigl( \sum_{D \in \frD_\infty^\eta} \psi_\infty^\eta(D)
  \Bigr)\, \prod_i \hat\caZ_{\infty,i}^\eta
\end{equation}
\begin{proposition}\label{Mayer2}
There exist constants $\be_6 \geq \be_5, c_6 > 0$ such that for any
$\be \geq l_0 \be_6$, $\eta \in \Om^{**}$, and volume $\La(N)$, $N \geq N^{**}(\eta)$, one has the bound
\begin{equation}
  \sum_{D \in \frD_\infty^\eta} |\psi_\infty^\eta(D)| \leq e^{-c_6 l_\infty}
\end{equation}
\end{proposition}

Gathering all expansion steps, we arrive at the final expression for the partition function $\caZ_\La^\eta$ in the form
\begin{equation}\label{eq: expansion}
\begin{split}
  \log \caZ^\eta_\La = &-E^\eta(\emptyset) +
  \sum_{C \in \frC_0^\eta} \phi_0^\eta(C)
  + \sum_{n \geq 1} \sum_{D \in \frD_n^\eta} \psi_n^\eta(D)
  + \sum_{D \in \frD_\infty^\eta} \psi_\infty^\eta(D)
\\
  &+\sum_i \log\hat\caZ_{\infty,i}^\eta
  +\sum_{n \geq 1} \sum_\al
  \log\hat\caZ_{n,\al}^\eta
\end{split}
\end{equation}
The terms collected on the first line contain the `vacuum' energy under the boundary condition $\eta$, together with the contributions of clusters of all orders. Recall that the latter allow for a uniform exponential upper bound. On the second line there are the partition functions of all $n$- and all corner aggregates. Although we can provide only rough upper bounds for these terms, a crucial property to be used is that the probability of an aggregate to occur is exponentially small in the size of its boundary, see Section~\ref{sec: classification}. In this sense, the above expansion is a natural generalization of the familiar `uniform' cluster expansion \cite{KoPr}.

\subsection{Estimates on the aggregate partition functions}

In expression~\eqref{eq: expansion} we do not attempt to perform any detailed expansion
of the aggregate's (log-)partition functions
$\hat\caZ_{n,\al}^\eta$ and $\hat\caZ_{\infty,i}^\eta$
via a series of local and exponentially damped terms. Instead, we follow the idea
that a locally ill-behaving boundary condition forces a partial coarse-graining represented above via the framework of aggregates of different orders. Although the detailed (cluster expansion-type) control within the aggregates is lost, we still can provide generic upper bounds on these partition functions. Notice a basic difference between $n$-aggregates and corner aggregates: The former contain only simple boundary contours the weights of which exponentially decay with the height of the contours. In some sense, the partition functions $\hat\caZ_{n,\al}^\eta$ can be compared with the partition function of a 1d interface to get an upper bound. On the other hand, the corner aggregates are ensembles of contours the weight of which obey no uniform exponential bound with the space extension of the contours, and allow possibly for a non-trivial `degeneracy of vacuum'. As a consequence, only rough (counting-type) estimates can be provided for the partition functions $\hat\caZ_{\infty,i}^\eta$.

\begin{lemma}\label{lem: hatcaZ-normal}
There are constants $c_7, c'_7 > 0$ ($c_7 \downarrow 0$ if $\be \to \infty$) such that for any $n$-aggregate $\caK_{n,\al}^\eta$, one has the bound
\begin{equation}
  \log\hat\caZ_{n,\al}^\eta \leq c_7 |\pa\caK_{n,\al}^\eta|
\end{equation}
For any corner aggregate $\caK_{\infty,i}^\eta$,
\begin{equation}
  \log\hat\caZ_{\infty,i}^\eta \leq c'_7 l_\infty^2
\end{equation}
\end{lemma}

\section{Asymptotic triviality of the constrained Gibbs measure $\nu_\La^\eta$}

As the first application of expansion~\eqref{eq: expansion} we prove that the weak limit of the constrained measure $\nu_\La^\eta$ coincides with the `+' phase Gibbs measure
$\mu^+$, finishing the first part of our program.

\begin{proposition}\label{prop: asym-triv}
There exists a constant $c > 0$ such that for any $\be \geq l_0\be_6$ (with the $\be_6$
the same as in Proposition~\ref{Mayer2}), any $\eta \in \Om^{**}$, and $X \subset \bbZ^2$ finite,
\begin{equation}
  \|\nu_{\La(N)}^\eta - \mu^+\|_X = O(e^{-cN})
\end{equation}
In particular, $\lim_{N\to\infty} \nu_{\La(N)}^\eta = \mu^+$, $\bsP$-a.s.
\end{proposition}
\begin{proof}
The idea of the proof is to express the expectation $\nu_{\La(N)}^\eta(f)$ of any local function $f$
as the sum of a convergent series by using the multi-scale scheme developed in the last section, and to compare the series with a standard cluster expansion for $\nu_{\La(N)}^{\eta \equiv +1}$. The difference between both series is given in terms of clusters both touching the boundary and the dependence set of $f$. Restricting only to the boundary conditions $\eta \in \Om^{**}$ and volumes $\La(N)$, $N \geq N^{**}(\eta)$ and using the exponential decay of the cluster weights, we prove the exponential convergence $\nu_{\La(N)}^\eta(f) \to \nu^+(f)$.

For notational simplicity, we only restrict to a special case and give a proof of the equality
\begin{equation}
  \lim_\La \nu_\La^\eta(\si_0 = -1) = \mu^+(\si_0 = -1)
\end{equation}
The general case goes along the same lines.

Assuming $\si \in \Om_\La^+$,
observe that $\si_0 = -1$ if and only if the set $\caD_\La(\si)$ contains an
odd number of contours $\Ga$ such that $0 \in \Int(\Ga)$. In an analogy with~\eqref{eq: polymer model}, we write the
$\nu_\La^\eta$-probability that $\si_0 = -1$ in the form
\begin{equation}\label{eq: expectation}
  \nu_\La^\eta(\si_0 = -1) = \frac{1}{\caZ_\La^\eta}
  \sum_{\De \sqsubset \La} \caZ_\La^\eta(\setminus\De)
  \prod_{\Ga \in \De} \rho^\eta(\Ga)
\end{equation}
where we have used the shorthand $\De \sqsubset \La$ for any compatible family of contours in $\La$ such that $\card(\De)$ is an odd integer and $0 \in \Int\Ga$
for every $\Ga \in \De$ . Furthermore, $\caZ_\La^\eta(\setminus\De)$ is the partition function
\begin{equation}
  \caZ_\La^{\eta}(\setminus\De) =
  \exp{(-E_\La^{\eta}(\emptyset))}\sum_{\partial\in\scD_{\La}(\setminus\De)}
  \prod_{\Ga\in\pa} \rho^{\eta}(\Ga)
\end{equation}
of a polymer model over the restricted ensemble
$\caK_\La(\setminus\De) \subset \caK_\La$ of all contours $\Ga$ such that
i) $\Ga \sim \De$, and ii) $0 \not\in \Int(\Ga)$. We can now repeat the same procedure as in the last sections, but with the contour ensemble $\caK_\La$ being replaced by
$\caK^\eta(\setminus\De)$. A crucial observation is that all contours from the set $\caK^\eta \setminus \caK^\eta(\setminus\De)$ are balanced, at least for all $\eta \in \Om^{**}$
and provided that the volume $\La(N)$ is large enough. Hence, the sets of unbalanced contours coincide for both contour ensembles $\caK^\eta$ and
$\caK^\eta(\setminus\De)$, hence, the same is true for the collections of
both $n$- and corner aggregates.
Finally, we compare the terms in the expansions for $\caZ_\La^\eta$ and $\caZ_\La^\eta(\setminus\De)$, and arrive at the formula
\begin{equation}
  \log\frac{\caZ_\La^\eta(\setminus\De)}{\caZ_\La^\eta}
  = -\sum_{C \in \frC_0^\eta \setminus \frC_0^\eta(\setminus\De)} \phi_0^\eta(C)
  - \sum_{n \geq 1} \sum_{D \in \frD_n^\eta \setminus \frD_n^\eta(\setminus\De)}
  \psi_n^\eta(D)
  - \sum_{D \in \frD_\infty^\eta \setminus \frD_\infty^\eta(\setminus\De)}
  \psi_\infty^\eta(D)
\end{equation}
where each of the three sums runs over all ($0$-, $n$-, or $\infty$-)clusters that are either incompatible with $\De$ or contain a contour $\Ga$, $0 \in \Int(\Ga)$. By construction, each $n$-, respectively $\infty$-cluster is further required to be incompatible with an $n$-, respectively corner aggregate, and since their weights are uniformly exponentially bounded by Propositions~\ref{Mayer}-\ref{Mayer2}, we get the uniform upper bound
\begin{equation}\label{eq: upperc}
  \sup_\La \Big|\log\frac{\caZ_\La^\eta(\setminus\De)}{\caZ_\La^\eta} \Big|
  \leq c|\De|
\end{equation}
with a constant $c$ large enough, as well as the existence of the limit
\begin{equation}\label{eq: limitc}
  \lim_\La \log\frac{\caZ_\La^\eta(\setminus\De)}{\caZ_\La^\eta}
  = -\sum^{\phantom{xxx},}_{C} \phi_0(C)
\end{equation}
where the sum runs over all finite $0$-clusters in $\bbZ^2$ that are either incompatible with $\De$ or contain a contour surrounding the origin. \\
Since every $\Ga \in \De$ surrounds the origin, it is necessarily balanced and  satisfies $\rho^\eta(\Ga) \leq \exp(-\frac{2\be}{l_0}|\Ga|)$. Combined with \eqref{eq: upperc}-\eqref{eq: limitc}, one easily checks that
\begin{equation}
  \lim_\La \nu_\La^\eta(\si_0 = -1) = \sum_{\De \sqsubset \bbZ^2}
  \exp\bigl( -\sum_{C}^{\phantom{xxx},} \phi_0(C) \bigr) \prod_{\Ga\in\De} \rho(\Ga)
\end{equation}
and the convergence is exponentially fast.
Obviously, the right-hand side coincides with the limit
$\lim_\La \mu_\La^{\eta\equiv+1}(\si_0 = -1) = \mu^+(\si_0 = -1)$, which finishes the proof.
\end{proof}

\section{Random free energy difference}\label{sec: prob-analysis}

In this section we analyze the limit behavior of the sequence of the random free energy differences
\begin{equation}
  F_\La^\eta = \log\caZ_\La^\eta - \log\caZ_\La^{-\eta}
\end{equation}
In order to show that the probability that $F_\La^\eta$ takes a value in a fixed finite interval is bounded as
$\caO(N^{-\frac{1}{2} + \al})$ with $\al > 0$, we can use the local central limit upper bound proven in Appendix~\ref{sec: LLT}, provided that a
Gaussian-type upper bound on the characteristic functions of the random variables $F_\La^\eta$ can be established. The basic idea is to prove the latter by employing
the sequential expansion for $\log\caZ_\La^\eta$ developed in section~\ref{sec: sequential} and by computing the characteristic functions in a neighborhood of the origin via a Mayer expansion. However, a technical problem arises here due to the high probability of the presence of corner aggregates. That is why we need to split our procedure in two steps that can be described as follows.

In the first step, we fix the boundary condition in the logarithmic neighborhood of the corners and consider the random free energy difference $F_\La^\eta$ conditioned on the fixed configurations. For this conditioned quantity a Gaussian upper bound on the characteristic function can be proven, implying a bound on the probability that the conditioned free energy difference $\bsP$-a.s.\ takes a value in a scaled interval $(aN^{\de},bN^{\de})$. This can be combined with a Borel-Cantelli argument to exclude all values in such an interval, at least $\bsP$-a.s.\ and for all but finitely many volumes from a sparse enough sequence of volumes.

In the second step, we consider the contribution to the free
energy difference coming from the corner aggregates. However,
their contribution to the free energy will be argued to be of a
smaller order when compared with the contribution of the
non-corner terms.

Note that we also include the $\infty$-clusters in the first step.
Because we have uniform bounds in $\eta$ for the $\infty$-cluster
weights, we are allowed to do so.

The free energy difference $F_\La^\eta$ can be computed by using
the sequential expansion~\eqref{eq: expansion}. For convenience,
we rearrange the terms in the expansion by introducing
\begin{equation}\label{eq: U}
\begin{split}
  U^\eta(B) &= \sum_n \sum_\al \log\hat\caZ_{n,\al}^\eta
  \funit_{\{\dom(\caK_{n,\al}^\eta) = B\}}
  + \sum_i \log\hat\caZ_{\infty,i}^\eta \funit_{\{\dom(\caK_{\infty,i}^\eta) = B\}}
\\
  &+ \sum_C \phi_0^\eta(C) \funit_{\{\dom(C) = B\}}
  + \sum_n \sum_D \psi_n^\eta(D) \funit_{\{\dom(D) = B\}}
\\
  &+ \sum_D \psi_\infty^\eta(D) \funit_{\{\dom(D) = B\}}
\end{split}
\end{equation}
for any set $B \subset \pa\La$.
Note that any function $U^\eta(B)$ only depends on the restriction of $\eta$ to the set
$\un{B}$.
Using the notation $\bar U^\eta(B) = U^\eta(B) - U^{-\eta}(B)$, the expansion for the free energy difference $F_\La^\eta$ reads, formally,
\begin{equation}
  F_\La^\eta = 2\be\sum_{x \in \un{\pa\La}} \eta_x
  + \sum_{B \subset \pa\La} \bar U^\eta(B)
\end{equation}
Obviously, no bulk contours contribute to $\bar U^\eta(B)$.
Using the notation
$\pa\La_{C,i} := \{y^*\in\pa\La:\,d[y^*,x^*_{C,i}] \leq 2l_\infty\}$ and $\pa\La_{C} := \cup_{i=1}^4 \pa\La_{C,i}$,
we consider the decomposition $F_\La^\eta = \tilde F_\La^\eta + \hat F_\La^\eta$, where
\begin{equation}
  \tilde F_\La^\eta = 2\be\sum_{x \in \un{\pa\La \setminus \pa\La_C}} \eta_x
  + \sum_{B \subset \pa\La \atop \dom(B) \not\subset \pa\La_C} \bar U^\eta(B)
\end{equation}
and
\begin{equation}
  \hat F_\La^\eta = 2\be\sum_{x \in \un{\pa\La_C}} \eta_x
  + \sum_{B \subset \pa\La \atop \dom(B) \subset \pa\La_C} \bar U^\eta(B)
\end{equation}
The first term, $\tilde F^\eta(B)$, can be analyzed by means of
the Mayer expansion of its characteristic function\\
\begin{equation}
\begin{split}
  \tilde\Psi_\La^\eta(t) &:= \bsE [\exp(it\tilde F_\La^\eta) \rel \eta_{\un{\pa\La_C}}]
  = \bsE \bigl[ \exp \bigl( 2it\be\sum_{x \in \un{\pa\La \setminus \pa\La_C}}
  \eta_x \bigr) \sum_{\caB} \prod_{B\in\caB} (e^{it\bar U^\eta(B)} - 1) \rel
  \eta_{\un{\pa\La_C}} \bigr]
\\
  &= [\Psi_0(t)]^{|\pa\La \setminus \pa\La_C|}
  \, \sum_{\caB} w_t(\caB \rel \eta_{\un{\pa\La_C}})
\label{eq: char-1-step}
\end{split}
\end{equation}
where we have assigned to any family $\caB$ of subsets of the boundary the weight
\begin{equation}\label{eq: char-weight}
\begin{split}
  w_t(\caB \rel \eta_{\un{\pa\La_C}})
  &= \frac{1}{[\Psi_0(t)]^{|\pa\La \setminus \pa\La_C|}}\,
  \bsE\bigl[ \exp(2it\be\sum_{x \in \un{\pa\La \setminus \pa\La_C}} \eta_x)
  \prod_{B\in\caB} (e^{it\bar U^\eta(B)} - 1) \rel \eta_{\un{\pa\La_C}} \bigr]
\\
  &\times \funit_{\{\forall B \in \caB:\,B \not\subset\pa\La_C\}}
\end{split}
\end{equation}
and have introduced the notation
\begin{equation}
  \Psi_0(t) = \bsE[ \exp(2it\be\eta_0) ] = \cos 2t\be
\end{equation}
Observing that
\begin{equation}
  w(\caB_1 \cup \caB_2 \rel \eta_{\un{\pa\La_C}}) =
  w(\caB_1 \rel \eta_{\un{\pa\La_C}})\, w(\caB_2 \rel \eta_{\un{\pa\La_C}})
\end{equation}
whenever $B_1 \cap B_2 = \emptyset$ for any $B_1 \in \caB_1$ and $B_2 \in \caB_2$, the last sum in equation~\eqref{eq: char-1-step} is a partition function of another polymer model and using the symbols $\frB, \frB_1, \ldots$ for the clusters in this model and
$w_t^T$ for the cluster weights, we get
\begin{equation}\label{eq: char-expansion}
  \tilde\Psi_\La^\eta(t) = [\Psi_0(t)]^{|\pa\La \setminus \pa\La_C|}
  \exp \bigl[ \sum_{\caB} w_t^T(\caB \rel \eta_{\un{\pa\La_C}}) \bigr]
\end{equation}
A crucial observation is that for any $\eta \in \Om^{**}$ and $\La(N)$, $N \geq N^{**}(\eta)$ no corner aggregate contributes to the weight $w_t(\caB)$ for any $\caB$. On the other hand, the partition function of any $n$-aggregate is balanced by a small probability of the aggregate to occur. Another observation is that every weight $w_t(\caB)$ is of order $\caO(t^2)$ due to the symmetry of the distribution $\bsP$. To see this explicitly, formula~\eqref{eq: char-weight} can be cast into a more symmetrized form,
\begin{equation}\label{eq: char-weight-sym}
\begin{split}
  w_t(\caB \rel \eta_{\un{\pa\La_C}})
  &= \frac{1}{[\Psi_0(t)]^{|\supp(\caB)|}}\,
  \bsE\Bigl[ T\Bigl\{t\bigl[2\be\sum_{x \in \un{\supp(\caB)}} \eta_x
  + \frac{1}{2} \sum_{B \in \caB} \bar U^\eta(B) \bigr]\Bigl\}
\\
  &\times \prod_{B\in\caB} 2i\sin \Bigl(\frac{t \bar U^\eta(B)}{2}\Bigr)
  \,\Bigl|\, \eta_{\un{\pa\La_C}} \Bigr]
\end{split}
\end{equation}
where $\supp(\caB):=\cup_{B \in \caB} B$ and
\begin{equation}
  T\{Y\} :=
  \begin{cases}
    i\sin Y  &  \text{if } \card(\caB) = 2k-1
  \\
    \cos Y  &  \text{if } \card(\caB) = 2k \qquad k \in \bbN
  \end{cases}
\end{equation}
In section~\ref{sec: char-upper-bound-proof} we give a proof of the following upper bound on the corresponding cluster weights:
\begin{lemma}\label{lem: difficult}
There exist constants $\be_8, l_0 > 0$\footnote{
     Recall that the construction of aggregates depends on the choice of $l_0$.}
such that for any $\be \geq \be_8 l_0$ there is $t_0 = t_0(\be) > 0$ for which the following is true. For any $\eta \in \Om^{**}$ and $\La = \La(N)$, $N \geq N^{**}(\eta)$, the inequality
\begin{equation}
  \sup_{x^* \in \pa\La \setminus \pa\La_C} \sum_{\caB:\, x^* \in \supp(\caB)}
  |w_t^T(\caB \rel \eta_{\un{\pa\La_C}})| \leq \frac{1}{2}\be^2 t^2
\end{equation}
is satisfied for all $|t| \leq t_0$.
\end{lemma}
With the help of the last lemma, it is easy to get an upper bound on
$\tilde \Psi_\La^\eta(t)$:
\begin{lemma}\label{lem: char-upper-bound}
Under the assumptions of Lemma~\ref{lem: difficult}, we have
\begin{equation}
  \tilde\Psi_\La^\eta(t) \leq \exp\Bigl(-\frac{1}{2}\be^2 t^2
  |\pa\La(N) \setminus \pa\La_C(N)|\Bigr)
\end{equation}
for all $|t| \leq t_0$, $\eta \in \Om^{**}$, and $N \geq N^{**}(\eta)$.
\end{lemma}
\begin{proof}
It immediately follows by combining Lemma~\ref{lem: difficult}, equation~\eqref{eq: char-expansion}, and the bound $\Psi_0(t) \leq \exp[-\be^2 t^2]$.
\end{proof}

For the corner part $\hat F_\La^\eta$ of the free energy difference we use the next immediate upper bound:
\begin{lemma}\label{lem: corner-contribution}
Given $\eta \in \Om^{**}$ and $\be \geq \be_6 l_0$, then
$\hat F_{\La(N)}^\eta = \caO(N^\de)$ for any $\de > 0$.
\end{lemma}
\begin{proof}
Using Proposition~\ref{Mayer2} and Lemma~\ref{lem: hatcaZ-normal}, we have
$\sum_{B \subset \pa\La_C} |\bar U^\eta(B)| = \caO(l_\infty^2)$ and the above claim immediately follows.
\end{proof}

\begin{proof}[Proof of Proposition~~\ref{prop: basic-est}]
Combining Lemma~\ref{lem: char-upper-bound} with Proposition~\ref{prop: LLT} in the appendix, we get
\begin{equation}
  \varlimsup_{N \to \infty} N^{\frac{1}{2} - \al}
  \bsP\bigl\{|\tilde F_{\La(N)}^\eta| \leq N^\al \tau \,\bigl|\,
  \eta_{\un{\pa\La_C}}\bigr\} < \infty
\end{equation}
for any $\al, \tau > 0$. By Lemma~\ref{lem: corner-contribution}, $\tilde F$
can be replaced with the full free energy difference $F$. As a consequence,
\begin{equation}
  \varlimsup_{N \to \infty} N^{\frac{1}{2} - \al}
  \bsP\{|F_{\La(N)}^\eta| \leq \tau\} < \infty
\end{equation}
and the proof is finished by applying Proposition~\ref{prop: asym-triv}.
\end{proof}

\section{Proofs}\label{sec: proofs}

In this section, we collect the proofs omitted throughout the main
text.

\subsection{Proof of Proposition~\ref{prop: prob-bound}}\label{sec: prob-bound-proof}

In order to get the claimed exponential upper bound on the probability for an $n$-aggregate to occur, we need to analyze the way how the aggregates are constructed in more detail. We start with an extension of Definition~\ref{def: decomposition}.
Throughout the section, a finite volume $\La = \La(N)$ is supposed to be fixed.
\begin{definition}
For every $n = 1,2,\ldots$, any maximal $L_n$-connected subset
$\De \subset \caK \setminus (\caK_0^\eta \cup \caK_1^\eta \cup \ldots\cup \caK_{n-1}^\eta)$ is called an $n$-pre-aggregate.
\end{definition}
Obviously, $n$-aggregates are exactly those $n$-pre-aggregates $\De$ that satisfy the condition $|\pa\De|\con \leq l_n$. Moreover, every $n$-pre-aggregate can equivalently be constructed inductively by gluing pre-aggregates of lower orders:
\begin{lemma}\label{lem: proof-decomposition}
Every $n$-pre-aggregate $\De_n$ is the union of a family of $(n-1)$-pre-aggregates,
$\De_n = \cup_\al \De_{n-1}^\al$. Moreover,
\begin{enumerate}
\item[i)]
Each $(n-1)$-pre-aggregate $\De_{n-1}^\al$ satisfies
$|\pa\De_{n-1}^\al|\con > l_{n-1}$,
\item[ii)]
The family $(\De_{n-1}^\al)_\al$ is $L_n$-connected.
\end{enumerate}
\end{lemma}
\begin{proof}
For $n=1$ the statement is trivial.\\ Assume that $n \geq 2$, and let
$\De_n$ be an $n$-pre-aggregate and $\Ga \in \De_n$ be a contour.
Then, there exists an $(n-1)$-pre-aggregate $\De_{n-1}^\al$ such that
$\Ga \in \De_{n-1}^\al$ (otherwise $\Ga$ would be an element of a
$k$-aggregate, $k \leq n-2$). Moreover, since
$\De_{n-1}^\al$ is not an $(n-1)$-aggregate by assumption, it satisfies
$|\pa\De_{n-1}^\al|\con > l_{n-1}$, proving i).

The claim ii) is obvious.
\end{proof}

\begin{lemma}\label{lem: proof-preliminary}
Let $\De$ by any family of unbalanced contours. Then,
\begin{enumerate}
\item[i)]
There exists a subset $\tilde\De \subset \De$ such that
\begin{enumerate}
\item[a)]
$\pa\tilde\De = \pa\De$,
\item[b)]
if $\Ga_1,\Ga_2,\Ga_3 \in \tilde\De$ are any three mutually different contours, then $\pa\Ga_1 \cap \pa\Ga_2 \cap \pa\Ga_3 = \emptyset$.
\end{enumerate}
\item[ii)]
The inequality
\begin{equation}\label{eq: LD-bound1}
  \sum_{x \in \un{\pa\De}} \eta_x < -\bigl( 1 - \frac{4}{l_0} \bigr) |\pa\De|
\end{equation}
holds true.
\end{enumerate}
\end{lemma}
\begin{proof}
i) Assume that $\Ga_1,\Ga_2,\Ga_3 \subset \De$ is a triple of mutually different contours such that
$\pa\Ga_1 \cap \pa\Ga_2 \cap \pa\Ga_3 \neq \emptyset$. Since
$\pa\Ga_i$, $i = 1,2,3$ are connected subsets of $\pa\La$, it is easy to realize that, up to a possible permutation of the index set $\{1,2,3\}$, one has
$\pa\Ga_1 \subset \pa\Ga_2 \cup \pa\Ga_3$. Hence,
$\pa(\De \setminus \{\Ga_1\}) = \pa\De$.
Since the set $\De$ is finite, a subset $\tilde\De \subset \De$ with
the claimed property is constructed by iterating the argument.

ii) Let $\tilde\De \subset \De$ be the same as in i). Then, using Lemma~\ref{lem: geom-balanced}, the inclusion-exclusion principle implies
\begin{equation}
\begin{split}
  \sum_{x \in \un{\pa\De}} \eta_x &= \sum_{\Ga \in \tilde\De}
  \sum_{x \in \un{\pa\Ga}} \eta_x
  - \sum_{(\Ga,\Ga') \subset \tilde\De}
  \sum_{x \in \un{\pa\Ga \cap \pa\Ga'}} \eta_x
\\
  &< -\bigl( 1-\frac{2}{l_0} \bigr) \sum_{\Ga \in \tilde\De} |\pa\Ga|
  + \sum_{(\Ga,\Ga') \subset \tilde\De} |\pa\Ga \cap \pa\Ga'|
\\
  &\leq -\bigl( 1-\frac{4}{l_0} \bigr) |\pa\De|
\end{split}
\end{equation}
\end{proof}

It remains to prove that one still gets a large deviation upper bound
by replacing the sum over the boundary sites $x \in \un{\pa\De}$ in
equation~\eqref{eq: LD-bound1} with the sum over all $x \in
\un{\cn(\pa\De)}$, provided that $\De$ is a pre-aggregate.
Technically, we need to exploit the basic feature of any
pre-aggregate $\De$ that the set $\cn(\pa\De) \setminus \pa\De$ is
not `too big'. A minor complication lies in the fact that the
boundary distance $d[\pa\ga,\pa\ga']$ is allowed to exceed the
contour distance $d[\ga,\ga']$. To overcome this difficulty, it is
useful to define
\begin{equation}
  \widetilde{\cn}(\pa\De) = \pa\De \cup \bigl\{x \in \cn(\pa\De);\,
  \forall \ga \in \De:\, d[x,\pa\ga] > \frac{|\pa\ga|}{l_0} \bigr\}
\end{equation}
for which the first equation in the proof of Lemma~\ref{lem: geom-balanced} implies the upper bound
\begin{equation}\label{eq: technical}
  |\pa\De|\con \leq (1+\frac{2}{l_0})|\widetilde{\cn}(\pa\De)|
\end{equation}

We are now ready to prove the following key estimate from which
Proposition~\ref{prop: prob-bound} immediately follows by using a
large deviation upper bound.
\begin{lemma}
Let $\De$ be an $n$-pre-aggregate, $n = 1,2,\ldots$ Then,
\begin{equation}\label{eq: overcome}
  \sum_{x \in \un{\cn(\pa\De)}} \eta_x \leq -\frac{1}{3} |\pa\De|\con
\end{equation}
uniformly in $n$.
\end{lemma}
\begin{proof}
We prove by induction in the order of the pre-aggregates the refined bound
\begin{equation}
  \sum_{x \in \un{\widetilde{\cn}(\pa\De)}} \eta_x \leq
  - \bigl( 1 - 3\sum_{i = 1}^{n} \frac{L_i}{l_{i-1}} \bigr) |\widetilde{\cn}(\pa\De)|
\end{equation}
for any $n$-pre-aggregate $\De$, from which the statement follows by
using the definition~\ref{assumptions scales} of length scales $l_n$
and $L_n$, and equation~\eqref{eq: technical}. Indeed, one obtains
then
\begin{equation}
\begin{split}
  \sum_{x\in\cn(\pa\De)} \eta_x &\leq  -\bigl( 1 - 3\sum_{i=1}^\infty \frac{L_i}{l_{i-1}} \bigr)
  |\widetilde{\cn}(\pa\De)| + |\cn(\pa\De) \setminus \widetilde{\cn}(\pa\De)|
\\
  &\leq -\bigl(\frac{1}{4} - \frac{2}{l_0}\bigr) \frac{|\pa\De|\con}{1+\frac{2}{l_0}}
  \leq -\frac{1}{3}|\pa\De|\con
\end{split}
\end{equation}

First, assume that $\De$ is a 1-pre-aggregate, and let
$\De = \cup_{i=1}^m A_i$ be the (unique) decomposition of $\De$ into disjoint subsets such that
$\pa\De = \cup_{i=1}^m \pa A_i$ is the decomposition of $\pa\De$ into maximal connected components. For convenience, we use the notation
$J_i := \pa A_i$. Considering furthermore the decomposition
$\widetilde{\cn}(\pa\De) \setminus \pa\De = \cup_{k=1}^{m-1} G_k$ into maximal connected components, the set $\widetilde{\cn}(\pa\De)$ can be finally written as the union
\begin{equation}\label{eq: boundary decomposition}
  \widetilde{\cn}(\pa\De) = \bigl( \cup_{i=1}^m J_i \bigr) \cup \bigl( \cup_{k=1}^{m-1} G_k \bigr)
\end{equation}
of disjoint connected subsets, which satisfy the inequalities
$|J_i| > l_0$ and $|G_k| \leq L_1$, for all $i,k = 1,2,\ldots$ Using Lemma~\ref{lem: proof-preliminary}, we have
$\sum_{x \in \un{J_i}} \eta_x \leq -\bigl( 1 - \frac{4}{l_0} \bigr) |J_i|$,
and since $\sum_{k=1}^{m-1} |G_k| \leq \frac{L_1}{l_0} \sum_{i=1}^m |J_i|$,
we finally get
\begin{equation}
\begin{split}
  \sum_{x \in \un{\widetilde{\cn}(\pa\De)}} \eta_x &= \sum_{i=1}^m \sum_{x \in \un{J_i}} \eta_x
  + \sum_{k=1}^{m-1} |G_k|
  \leq - \bigl( 1 - \frac{L_1 + 4}{l_0} \bigr)
  \frac{|\widetilde{\cn}(\pa\De)|}{1 + \frac{L_1}{l_0}}
\\
  &\leq -\bigl( 1 - \frac{3L_1}{l_0} \bigr) |\widetilde{\cn}(\pa\De)|
\end{split}
\end{equation}
provided that, say, $L_1 \geq 4$. \\
Next, we will prove the statement for an arbitrary $n$-pre-aggregate $\De$. By Lemma~\ref{lem: proof-decomposition}, $\De$ is the union of a family of $(n-1)$-pre-aggregates, $\De = \cup_i \De_{n-1}^i$.
In order to generalize our strategy used in the $n=1$ case, we consider the (possibly disconnected) boundary sets
$J_i = \widetilde{\cn}(\pa\De_{n-1}^i)$, and the family of connected sets $(G_i)_{i=1,2,\ldots}$ defined as the maximal connected components of the set $\cn(\pa\De) \setminus \cup_i \cn(\De_{n-1}^i)$. Note that
$\#\{G_i\} = \#\{J_i\} - 1$ and the identity
$\widetilde{\cn}(\De) = (\cup_i J_i) \cup (\cup_i G_i)$. Hence, by using the induction hypothesis,
\begin{equation}
\begin{split}
  \sum_{x \in \un{\widetilde{\cn}(\pa\De)}} \eta_x &= \sum_{i=1}^m \sum_{x \in \un{J_i}} \eta_x
  + \sum_{k=1}^{m-1} |G_k|
  \leq \Bigl[ -\bigl(1 - 3\sum_{i=1}^{n-1}\frac{L_i}{l_{i-1}} \bigr) + \frac{L_n}{l_{n-1}}
  \Bigr] \frac{|\widetilde{\cn}(\pa\De)|}{1+\frac{L_n}{l_{n-1}}}
\\
  &\leq -\bigl( 1 - 3\sum_{i=1}^n \frac{L_i}{l_{i-1}} \bigr) |\widetilde{\cn}(\pa\De)|
\end{split}
\end{equation}
as required.
\end{proof}

\subsection{Proof of Proposition~\ref{Mayer}}\label{sec: Mayer-proof}

The proof goes by induction in the order of aggregates.\\ \\
{\bf The case $n=1$.}\\
As the initial step we bound the sums over 1-clusters in
$\frD_1^\eta$. Recall that the 1-clusters consist of 0-clusters which connect 1-aggregates $\caK_{1,\al}^\eta$. Throughout this section we use the shorthand
$\tilde\be := \frac{\be}{l_0}$.\\\\
From Proposition \ref{le: clusters step zero} we know that for
any integer $r_0$,
\[\sum_{C\in\frC_0^\eta:~|C|\geq r_0\atop C\ni x} |\phi_0^\eta(C)|\exp{(\bet(2-(1/8))|C|)}\leq 1\]
which implies
 \beq\label{es: 0-cluster}
\sum_{C\in\frC_0^\eta:~|C|\geq r_0\atop C\ni x}
|\phi_0^\eta(C)|\exp{(2\bet(1-(1/8))|C|)}\leq \exp{(-\bet r_0/8)}
\eeq
We split the procedure into four steps as follows.\\\\
{\bf Part 1. } For any 1-cluster in $\frD_1^\eta$, none of its
0-clusters contributes to the dressed weight of a $1$-aggregate.
Hence, all these 0-clusters have at least
size $L_1$. Moreover, they are incompatible with a
1-aggregate $\caK_{1,\al}^\eta$. Using Lemma~\ref{lem: geom-balanced}
and choosing $r_0=L_1$ in \eqref{es: 0-cluster}, this results in the inequality
\beq\label{C<->K1al} \sum_{C\in\frC_0^\eta\atop
C\not\sim\caK_{1,\al}^\eta}
|\phi_0^\eta(C)|\exp{(2\bet(1-(1/8))|C|)}\leq l_1^2
\exp{\bigl[-(\bet L_1)/8\bigr]}\leq 2^{-2} \eeq {\bf Part 2. }
In order to prove the convergence of the cluster expansion
resulting from the Mayer expansion, we apply Proposition~\ref{prop: cluster model}.
As our initial estimate, we get, using \eqref{C<->K1al} and since
\[C\stackrel{1}{\not\lra} C'~\Leftrightarrow~\exists\al\mbox{ such that
}C,C'\not\sim\caK_{1,\al}^\eta\]
the inequality
\begin{equation}
\begin{split}
  \sum_{C\in\frC_0^\eta\atop C\stackrel{1}{\lra}C_0}
  &|\phi_0^\eta(C)|\exp{(2\bet(1-(1/8))|C|)}
\\
  &\leq\sum_{\caK_{1,\al}^\eta:~\caK_{1,\al}^\eta\inc
  C_0}\sum_{C\in\frC_0^\eta\atop
  C\inc\caK_{1,\al}^\eta}\exp{(2\bet(1-(1/8))|C|)}|\phi_0^\eta(C)|
\\
  &\leq 2^{-2}\#\{\caK_{1,\al}^\eta\inc
  C_0\}
\end{split}
\end{equation}\\\\
{\bf Part 3. } Using Lemma \ref{lem: n-cluster}, the weight of any set of
0-clusters appearing in the Mayer expansion is bounded as
\[|w_1^\eta(\caC)|\leq\prod_{C\in\caC}(e^{|\phi_0^\eta(C)|}-1)
\leq \prod_{C\in \caC} 2|\phi_0^\eta(C)|\]
Hence, by using Proposition~\ref{prop: cluster model}, we obtain the
bound \beq\label{G<->C0} \sum_{C_1\stackrel{1}{\not\lra} C_0\atop
C_1\in\frD_1^\eta}|\psi_1^\eta(C_1)|\exp{[(2\bet(1-(1/8))-1/2)|C_1|]}
\leq 2^{-1}\#\{\caK_{1,\al}^\eta\inc C_0\}\eeq
Taking now $C_0\in \frC_0^\eta$ such that $\caK_{1,\al}^\eta$ is the only
$1$-aggregate satisfying $C_0 \nsim \caK_1^\al$,
inequality~\eqref{G<->C0} yields
\beq\label{G<->Kla1} \sum_{C_1\inc\caK_{1,\al}^\eta\atop
C_1\in\frD_1^\eta}
|\psi_1^\eta(C_1)|\exp{[(2\bet(1-(1/8))-1/2)|C_1| ]}\leq 2^{-1}
  \eeq
{\bf Part 4. } In order to bound the sum over all 1-clusters $C_1 \in \frD_1^\eta$
such that $C_1\ni x$ and $|C_1|\geq r_1$, we use that
$|C_1|\geq L_1$ and write
\begin{equation}
\begin{split}
\sum_{C_1\ni x,~|C_1|\geq
r_1\atop C_1\in\frD_1^\eta}
&|\psi_1^\eta(C_1)|\exp{[(2\bet(1-(1+1/2)/8)-1/2)|C_1|]}
\\
\label{ind Mayer in between}
&\leq\sum_{\caK_{1,\al}^\eta}\sum_{C_1\inc\caK_{1,\al}^\eta:~C_1\ni
x\atop|C_1|\geq r_1,~C_1\in\frD_1^\eta}
|\psi_1^\eta(C_1)|\exp{[(2\bet(1-(1+1/2)/8)-1/2)|C_1|]}
\end{split}
\end{equation}
Substituting \eqref{G<->Kla1}, we obtain
\begin{equation}
\begin{split}
\eqref{ind Mayer in between} &\leq
\sum_{\caK_{1,\al}^\eta}2^{-1}\exp{(-(\bet/8)\cdot\max{[d(\caK_{1,\al}^\eta,x),r_1]})}
\\
&\leq \sum_{R=0}^\infty\sum_{\caK_{1,\al}^\eta:~d(\caK_{1,\al}^\eta,x)=R}
2^{-1}\exp{(-\ep\bet\cdot\max{[R,r_1]})}
\end{split}
\end{equation}
The last sum can be estimated by a partial integration and we finally get
\[\eqref{ind Mayer in between}\leq\exp{(-\bet r_1/8)}\left[r_1^2+\frac{16r_1}{\bet}+2\right]\leq
2^{-1}\cdot4r_1^2\exp{(-\bet r_1/8)}\]
where we have used that $r_1\geq L_1$ and that $L_1$ is large enough.

\noindent\\\\
{\bf Induction step.}\\
The induction hypothesis reads
\begin{equation}
\begin{split}
  \sum_{C_i\ni x:~|C_i|\geq r_i\atop C_i\in\frD_i^\eta}
  &|\psi_i^\eta(C_i)|\exp{  \left[\Bigl(2\bet(1-\sum_{j=0}^{i+1}
  (1/2)^j/8)-\sum_{j=1}^i(1/2)^j\Bigr)|C_i|\right]}
\\ \label{graph summation until n-1}
  &\leq 4\cdot 2^{-i}
  r_i^2\exp{(-\bet(1/2)^{i+1}r_i/8)}
\end{split}
\end{equation}
for any $1\leq i\leq n-1$.\\\\
{\bf Part 1. }As in part 1 of the $n=1$ case, we want to
prove first that \beq\label{C<->Klan}
\sum_{C\in\frC_{n-1}^\eta\atop C\inc\caK_{n,\al}^\eta}
|\phi_{n-1}^\eta(C)|\exp{\left[
\Bigl(2\bet(1-\sum_{j=0}^{n}(1/2)^j/8)-\sum_{j=1}^{n-1}(1/2)^j\Bigr)|C|\right]}\leq
2^{-n-1}\eeq
Recalling Definition~\eqref{eq: definition n-cluster weights} for
$\phi_{n-1}^\eta(C)$, we know that $\phi_{n-1}^\eta(C)=\psi_j^\eta(C)$ for any
$C\in\frD_j^\eta$. Hence, using \eqref{graph summation until n-1} with $r_i=L_n$,  we write
\begin{equation}
\begin{split}
\eqref{C<->Klan}&\leq l^2_n
\sum_{i=1}^{n-1}\sum_{C_i\ni x:~|C_i|\geq L_n\atop
C_i\in\frD_i^\eta}
|\psi_i^\eta(C_i)|\exp{\left[\Bigl(2\bet(1-\sum_{j=0}^n(1/2)^j/8)-
\sum_{j=1}^{n-1}(1/2)^j\Bigr)|C_i|\right]}
\\
  &\leq 4l_n^2L_n^2
  \sum_{i=0}^{n-1}2^{-i}\exp{
  \left[\Bigl(-\bet\sum_{j=i+1}^n(1/2)^j/4-\sum_{j=i+1}^{n-1}(1/2)^j\Bigr)L_n\right]}
\\
  &\leq 2^{-n-1}\cdot 32l_n^2L_n^2\exp{\left[-\bet(1/2)^n L_n/4\right]}\leq 2^{-n-1}
\end{split}
\end{equation}
where we have used that $l_n=\exp{(L_n / 2^n)}$ and $\bet$ is
large enough. This proves inequality~\eqref{C<->Klan}.\\\\
{\bf Part 2. } Similarly as in the $n=1$ case, we prove by using
\eqref{C<->Klan} the inequality
\[\sum_{C\stackrel{n}{\not\lra} C_0\atop C\in\frC_{n-1}^\eta}
|\phi_{n-1}^\eta(C)|\exp{\left[
\Bigl(2\bet(1-\sum_{j=0}^{n}(1/2)^j/8)-\sum_{j=1}^{n-1}(1/2)^j
\Bigr)|C|\right]}\leq 2^{-n-1}\#\{\caK_{n,\al}^\eta\inc C_0\}\]
\\\\
{\bf Part 3. }
By construction, any $n$-cluster $C_n\in\frD_n^\eta$ consists of a family of
$0$-clusters $C_0 \in \frC_0^\eta$ and $i$-clusters $C_1 \in \frD_0^\eta$,
$0 \leq i \leq n-1$, which are all incompatible with $\caK_n^\eta$.
Using Lemma~\ref{lem: n-cluster} again, we have the upper bound
\[|w_n^\eta(C_n)|\leq\prod_{i=0}^{n-1}
\prod_{C\in C_n \cap \frD_i^\eta}2|\psi_i^\eta(C)|\]
where we have identified $\psi_0^\eta(.)\equiv\phi_0^\eta(.)$ and
$\frD_0^\eta\equiv\frC_0^\eta$. Applying Proposition \ref{prop: cluster model} with
$z(C)=2|\psi_i^\eta(C)|$ then gives
\[\sum_{C_n\stackrel{n}{\not\lra} C_0\atop C_n\in\frD_n^\eta}
|\psi_n^\eta(C_n)|\exp{\left[\Bigl(2\bet(1-\sum_{j=0}^{n}(1/2)^j/8)-\sum_{j=1}^n(1/2)^j
\Bigr)|C|\right]}\leq 2^{-n} \#\{\caK_{n,\al}^\eta\inc C_0\}\]
Taking again $C_0\in \frC_0^\eta$ such that
$\caK_{n,\al}^\eta\inc C_0$ implies the inequality
\begin{equation}\label{G<->Klan}
\sum_{C_n\inc\caK_{n,\al}^\eta\atop C_n\in\frD_n^\eta}
|\psi_n^\eta(C_n)|\exp{
\left[\Bigl(2\bet(1-\sum_{j=0}^{n}(1/2)^j/8)-\sum_{j=1}^n(1/2)^j\Bigr)|C_n|\right]}
\leq 2^{-n}
\end{equation}
\\\\
{\bf Part 4. } Repeating the argument for the $n=1$ case, we obtain the inequality
\begin{equation}
\begin{split}
\sum_{C_n\ni x,~|C_n|\geq r_n\atop C_n\in\frD_n^\eta}
&|\psi_n^\eta(C_n)|\exp{\left[\Bigl(2\bet(1-\sum_{j=0}^{n+1}(1/2)^j/8)
-\sum_{j=1}^n(1/2)^j\Bigr)|C_n|\right]}
\\
&\leq 2^{-n}\cdot 4r_n^2\exp{(-(1/2)^{n+1}\bet r_n/8)}
\end{split}
\end{equation}
Using that $r_n\geq L_n$ for any $C_n\in\frD_n^\eta$ and choosing $r_n=L_n$ proves the proposition for the weights $\psi_n^\eta$, $n = 1,2,\ldots$. \\

Equation~\eqref{eq: definition n-cluster weights} reads that  $\phi_n^\eta(C)=\psi_j^\eta(C)$
whenever $C\in\frD_j^\eta$ and $j \leq n$. Using further that
$\frC_n^\eta=\frC_0^\eta\cup\frD_1^\eta\cup\cdots\cup\frD_n^\eta$ and summing up
the cluster weights of the clusters of all orders yields
inequality~\eqref{eq: multi-scale-bound-phi}, which finishes the
proof.

\subsection{Proof of Proposition~\ref{Mayer2}}

Let $n_0$ be the same as in Section~\ref{sec: corners}.
Due to the second part of Proposition~\ref{Mayer},
\[\sup_x\sum_{C\ni x,~|C|\geq r_0\atop C\in\frC_{n_0}^\eta}
|\phi_{n_0}^\eta(C)|\exp{\left[(\be/4l_0)\right]|C|} \leq
2\exp{(-(3\be/4l_0)r_0)}\] According to the definition of the
corner-aggregates, we have
\[\sum_{C\inc\caK_{\infty,i}\atop C\in\frC_{n_0}^\eta}
|\phi_{n_0}^\eta(C)|\exp{\left[(\be/4l_0)|C|\right]} \leq
2l_\infty^2\exp{(-l_\infty(3\be/2l_0))}\leq
2^{-3}\exp{(-l_\infty(\be/l_0))}\]
Applying Proposition~\ref{prop: cluster model}, we obtain
\[\sum_{C\inc\caK_{\infty,i}\atop C\in\frD_{\infty}^\eta}
|\psi_\infty^\eta(C)|\exp{\left[(\be/4l_0)\right]|C|}
\leq2^{-2}\exp{(-l_\infty(\be/l_0))}\] which implies
\[\sum_{D\in\frD_{\infty}^\eta}|\psi_\infty^\eta(D)|\leq\exp{(-(3\be/2l_0)l_\infty)}\]

\subsection{Proof of Lemma~\ref{lem: hatcaZ-normal}}

Let $\eta \in \Om^{**}$ and $\caK_{n,\al}^\eta$ be an
$n$-aggregate, $n = 1,2,\ldots$. Recall that
\begin{equation}
  \hat\caZ_{n,\al}^\eta = \sum_{\pa \in\scD_{n,\al}^\eta}
  \hat\rho^\eta(\pa)
\end{equation}
where
\begin{equation}
  \hat\rho^\eta(\pa) = \prod_{\Ga \in \pa} \rho^\eta(\Ga)\,
  \exp\Bigl( -\sum_{C\in \frC_{n-1}^\eta\atop C\not\sim\pa;~|C|<L_n} \phi_{n-1}^\eta(C) \Bigr)
\end{equation}
Using the $\eta$-uniform bounds
\begin{equation}
  \rho^\eta(\Ga) \leq \exp[ -2\be(|\Ga| - |\pa\Ga|)]
\end{equation}
and
\begin{equation}
 \sup_{x^\star}\sum_{C\in \frC_{n-1}^\eta\atop x^\star\in C} |\phi_{n-1}^\eta(C)|
  \leq \exp \bigl[-\frac{3\be}{l_0}\bigr]
\end{equation}
for all $n=1,2,\ldots$, one can subsequently write (for
simplicity, we use the shorthand $\ve =
\exp\bigl[-\frac{3\be}{l_0}]$ below):
\begin{equation}\label{eq: est-hatcaZ}
\begin{split}
  \hat\caZ_{n,\al}^\eta &\leq \sum_{\pa \in\scD_{n,\al}^\eta}
  \prod_{\Ga \in \pa} \exp [-(2\be - \ve)|\Ga| + 2\be|\pa\Ga|]
\\
  &\leq e^{(\ve + 4e^{-2\be + \ve})|\pa\caK^\eta_{n,\al}|} \sum_{\pa
  \in\scD_{n,\al}^\eta}
  \prod_{\Ga \in \pa} \exp [-(2\be - \ve)|\Ga| + (2\be-\ve - 4e^{-2\be + \ve}))|\pa\Ga|]
\\
  &\leq e^{(\ve + 4e^{-2\be + \ve})|\pa\caK^\eta_{n,\al}|} \sum_{\pa \in\scD^\eta_{n,\al}}
  \prod_{\Ga \in \pa} \prod_{\ga \in \Ga}
  \exp [-(2\be - \ve)|\ga| + (2\be - \ve -4e^{-2\be + \ve})|\pa\ga|]
\\
  &\leq e^{(\ve + 4e^{-2\be + \ve})|\pa\caK^\eta_{n,\al}|} \Bigl\{
  1 + \sum_{\ga \ni p} \exp [-(2\be - \ve)|\ga| + (2\be - \ve - 4e^{-2\be + \ve}))|\pa\ga|]
  \Bigr\}^{|\pa\caK^\eta_{n,\al}|}
\end{split}
\end{equation}
where the last sum runs over all pre-contours (= connected
components of contours) such that a fixed dual bond
$p=\pair{x,y}^\star,~d(x,\La^c)=d(y,\La^c)=1$ is an element of
$\ga$ and it is the leftmost bond with these properties, w.r.t.\ a
fixed orientation on the boundary. To estimate this sum, we
associate with each pre-contour $\ga$ a path (= sequence of bonds;
not necessarily unique) starting at $p$. Every such a path
consists of steps choosing from three of in total four possible
directions. One easily realizes that, for every such a path, the
total number of steps to the right is bounded from below by
$|\pa\ga|$. Hence, the last sum in \eqref{eq: est-hatcaZ} is
upper-bounded via the summation over all paths started at $p$, so
that to each step going to the right (respectively to the
left/up/down) one assigns the weight $e^{-4e^{-2\be+\ve}}$
(respectively $e^{-2\be + \ve}$), which yields
\begin{equation}
\begin{split}
  \sum_{\ga \ni p} &\exp [-(2\be - \ve)|\ga| + (2\be - \ve - 4e^{-2\be + \ve}))|\pa\ga|]
\\
  &\leq e^{-2\be + \ve} \sum_{n=1}^{\infty} \bigl( 2e^{-2\be+\ve} +
  e^{-4e^{-2\be + \ve}} \bigr)^n \leq 2e^{-2\be + \ve}
\end{split}
\end{equation}
All in all, one obtains
\begin{equation}
  \hat\caZ^\eta_{n,\al} \leq
  e^{(\ve + 6e^{-2\be+\ve})|\pa\caK^\eta_{n,\al}|}
\end{equation}
proving the first part of the statement.

The proof of the second part is trivial by counting the number of all
configurations in the square volume with side $2l_\infty$. Note that
the latter contains all contours $\Ga \in \pa$ for any configuration
$\pa\in \scD_{\infty,i}^\eta$ and that the weights of all clusters
renormalizing the contour weights are sumable due to
Proposition~\ref{Mayer}.

\subsection{Proof of Lemma~\ref{lem: difficult}}\label{sec: char-upper-bound-proof}

Due to Proposition~\ref{prop: cluster model}, it is enough to show that the inequality
\begin{equation}\label{eq: char to prove}
  \sum_{\caB:\,x^* \in \supp(\caB)}
  |w_t(\caB \rel \eta_{\un{\pa\La_C}})|\,
  \exp\bigl( \frac{1}{2}\be^2 t^2 |\supp(\caB)| \bigr)  \leq
  \frac{1}{2}\be^2 t^2
\end{equation}
holds true for all $|t| \leq t_0$, with a constant $t_0 > 0$. Remark that the RHS of the last equation is not optimal and can be improved, as obvious from the computation below.

In order to prove \eqref{eq: char to prove}, we use the symmetric representation~\eqref{eq: char-weight-sym} of
the weight $w_t(\caB \rel \eta_{\un{\pa\La_C}})$, the lower
bound $\Psi_0(t) \geq e^{-\al}$ which is true for any $\al > 0$ provided that
$|t| \leq t_1(\al)$ with a constant $t_1(\al) > 0$, and the estimate
\begin{equation}
\begin{split}
  \Bigl|T\Bigl\{t\bigl[&2\be\sum_{x \in \un{\supp(\caB)}} \eta_x
  + \frac{1}{2} \sum_{B \in \caB} \bar U^\eta(B) \bigr]\Bigl\}\Bigr|
\\
  &\leq \begin{cases}
         t\Bigl[2\be\sum_{x \in B}|\eta_x| + \frac{1}{2}|\bar U^\eta(B)|\Bigr]
         &   \text{for } \caB = \{B\}
       \\
         1  &  \text{otherwise}
       \end{cases}
\end{split}
\end{equation}
which will be enough in order to get the $t^2$ factor in what follows.
Using Proposition~\ref{Mayer} and Lemma~\ref{lem: hatcaZ-normal}, we get a uniform upper bound $|\bar U^\eta(B)| \leq c|B|$ with a constant $c > 0$ such that $c\downarrow 0$ for $\be \uparrow \infty$. Hence, in the case $\caB = \{B\}$ we have
\begin{equation}\label{eq: char one set}
\begin{split}
  |w_t(\caB &= \{B\})\rel \eta_{\un{\pa\La_C}}|
\\
  &\leq e^{\al|B|}t^2 \bsE \Bigl[ \bigl(
  2\be\sum_{x \in B} |\eta_x| + \frac{1}{2} \sum_{B \in \caB}
  |\bar U^\eta(B)| \bigr) \prod_{B \in \caB} |\bar U^\eta(B)|
  \,\Big|\, \eta_{\un{\pa\La_C}}\Bigr]
\\
  &\leq  e^{\al|B|} t^2 (2\be + \frac{c}{2})|B|
  \,\bsE \Bigl[ |\bar U^\eta(B)| \,\Bigl|\, \eta_{\un{\pa\La_C}} \Bigr]
\end{split}
\end{equation}
Note that the above uniform upper bound on $|\bar U^\eta(B)|$ is not sufficient to get a sensible estimate on the conditional expectation. However, a more detailed upper bound can be obtained. Without loss of generality, we can assume that $B \cap \pa\La_C = \emptyset$, so that the conditioning on $\eta_{\un{\pa\La_C}}$ can be omitted. First, assume there is an aggregate\footnote{For simplicity, we suppress the subscript $n$ here.} $\caK_\al^\eta$ such that $\dom(\caK_{\al}^\eta) = B$. Then,
Lemma~\ref{lem: hatcaZ-normal} gives the estimate
$\log\hat\caZ_{\al}^\eta \leq c_7 |B|$ and, since $|\pa\caK_{\al}^\eta| \geq |B|/2$, Proposition~\ref{prop: prob-bound} reads that the probability of such an event is bounded by $\exp(-\frac{c_5}{2}|B|)$. Second, assume there is a family of aggregates
$(\caK_{\al_i}^\eta)_i$ (of possibly different orders) such that $D^\eta := \cup_i \dom(\caK_{\al_i}^\eta) \subset B$.
Then, any cluster $C$ such that $\dom(C) = B$ has the length
$|C| \geq |B \setminus D^\eta|$
and Proposition~\ref{Mayer} gives the estimate
\[
  \sum_{C \in \cup_n \frC_n^\eta \atop \dom(C) = B} |\phi_n^\eta|
  \leq \exp\bigl(-\frac{\be}{2l_0}|B \setminus D^\eta|\Bigr)
\]
Moreover, the probability that $D^\eta = D$ for a fixed set $D$ is bounded by $e^{-\frac{c_5}{2}|D|}$. Note, however, that the above two scenarios are possible only provided that $|B| \geq l_1$, otherwise we only get a contribution from $0$-clusters, the sum of which is bounded by $e^{-\frac{\be}{2l_0}|B|}$. All in all, we obtain
\begin{equation}\label{eq: crucial}
\begin{split}
  \bsE \Bigl[ | U^\eta(B)| \Bigr]
  &\leq c_7|B|\,e^{-\frac{c_5}{2}|B|}\funit_{|B| \geq l_1} + e^{-\frac{\be}{2l_0}|B|}
  + \funit_{|B| \geq l_1} \sum_{D \subset B}
  e^{-\frac{c_5}{2}|D| - \frac{\be}{2l_0}|B \setminus D|}
\\
  &\leq e^{-\frac{\be}{2l_0}|B|} + \funit_{|B| \geq l_1} (c_7 + 1)|B|\,
  e^{-\frac{c_5}{4}|B|}
\end{split}
\end{equation}
provided that $\be / l_0$ is large enough. Recall that $c_5$ does not depend on $l_0$, which means that the latter can be adjusted as large as necessary. Using the same argument for $U^{-\eta}(B)$ and substituting
\eqref{eq: crucial} into \eqref{eq: char one set}, we get
\begin{equation}
\begin{split}
  \sum_{B \ni x \atop B \subset \pa\La}
  &|w_t(\caB = \{B\})\rel \eta_{\un{\pa\La_C}})|\, e^{\tau|B|}
  \leq 2\cdot t^2 (2\be + \frac{c}{2}) \sum_{B \ni x \atop B \subset \pa\La}
  |B|\,e^{(\tau + \al)|B|}\,
\\
  &\times \Bigl[ e^{-\frac{\be}{2l_0}|B|} + \funit_{|B| \geq l_1} (c_7 + 1)|B|\,
  e^{-\frac{c_5}{4}|B|} \Bigr] \leq \tau' \be t^2
\end{split}
\end{equation}
which is true for any $\tau' > 0$ provided that $\tau$ and $\al$ are chosen sufficiently small and $l_0$ (and hence $l_1$) sufficiently large. This argument can easily be generalized by taking into account all collections $\caB$, $\card(\caB) > 1$.
Hence, the proof of \eqref{eq: char to prove} is completed by choosing $\tau = \frac{1}{2}\be^2 t^2$, under the condition $|t| \leq t_0$ with $t_0 = t_0(\be)$ being small enough.

\section{Concluding remarks and some open questions}

Our result that a typical boundary condition (w.r.t.\ a symmetric
distribution)  suppresses both mixed and interface states explains
why these states are typically not observed in experimental
situations without a special preparation. To a certain extent it
justifies the standard interpretation of extremal invariant Gibbs
measures as pure phases.

Although this result, which finally solves the question raised in
\cite{NeSt97}, is only about the 2-dimensional Ising ferromagnet, and
thus seemingly of limited interest, it is our opinion that the
perturbation approach developed in the paper is actually very robust
(compare \cite{GBG}). As we have observed at various points in the
paper, there seems to be no barrier except some technical ones to
extend the analysis to the Ising model with random boundary
conditions in higher dimensions. In fact, there might be extensions
of our approach into various different directions. In particular,
both the random distribution of the boundary terms and the phase
transition itself could lack the plus-minus symmetry, and one might
also consider a more general Pirogov-Sinai set-up in which the number
of extremal Gibbs measures could be larger than two. Another possible
extension could be to finite-range Hopfield-type models, in which
periodic or fixed boundary conditions lack a coherence property with
respect to the possible Gibbs measures, and thus are expected to
behave as random ones~\cite{NSW}. Actually, our result can be
translated in terms of the Mattis (= single-pattern Hopfield) model
with fixed boundary conditions, proving the chaotic size-dependence
there.

More generally, in principle the phenomenon of the exclusion of
interface states for typical boundary conditions might well be of
relevance for spin glass models of Edwards-Anderson type, which has
indeed been one of our main motivations. Our result illustrates in a
simple way how the Newman-Stein metastate program, designed for the
models exhibiting the chaotic size-dependence, can be realized. The
number of states, as well as the number of ``physically relevant''
states for short-range spin glasses has been an issue of contention
for a long time. In this paper, we have provided a very precise
distinction between the set of all Gibbs states, the set of all
extremal Gibbs measures, and the set of ``typically visible'' ones,
without restricting \emph{a priori} to the states with a particular
symmetry. We hope the provided criterion might prove useful in a more
general context.

We mention that the restriction to sparse enough sequences of volumes is
essential to obtain almost sure results. Actually, for a regular sequence of
volumes, we expect all mixtures (in dimension three all translation-invariant
Gibbs measures) to be almost sure limit points, although in a null-recurrent
way. This still would mean that the metastate would not be affected, and
that it would be concentrated on the plus and minus measures. See also the
discussion in \cite{EMN}. However, proving this conjecture goes beyond the
presented technique and remains an open question.

\section*{Acknowledgements}

This research was supported by het Samenwerkingsverband Mathematische Fysica
FOM-GBE.

\appendix

\section{Cluster models}\label{ap: cluster models}

In this section we present a variant of the familiar result on the
convergence of the cluster expansion for polymer models, which proves
useful in the cases when the summation over polymers becomes
difficult because of their high geometrical complexity. Such a
situation arises, for example, in the applications of the cluster
expansion to the study of the convergence of high-temperature (Mayer)
series in lattice models with an infinite-range potential. Since the
Mayer expansion techniques are by no means restricted to the
high-temperature regimes (note e.g.\ its application in the RG
schemes for low-temperature contour models), the result below can be
applied in a wide class of problems under a perturbation framework.
In our context, we use the result to provide upper bounds on the
weights $\psi_n^\eta$ of $n$-clusters, see Section~\ref{sec:
Mayer-proof}.

We consider an abstract cluster model defined as follows. Let $G =
(S,\nsim)$ be a finite or countable non-oriented graph and call its
vertices \emph{polymers}. Any two polymers
$X \nsim Y$ are called \emph{incompatible}, otherwise they are \emph{compatible}, $X \sim Y$. By convention, we add the relations $X \nsim X$ for all $X \in S$. Any non-empty finite set $\De \subset S$ is called a \emph{cluster} whenever there exists no decomposition
$\De = \De_1 \cup \De_2$ such that $\De_1$ and $\De_2$ are non-empty disjoint sets of polymers and $\De_1 \sim \De_2$, where the latter means that $X \sim Y$ for all $X \in \De_1$ and $Y \in \De_2$. Let $\caP(S)$ denote the set of all finite subsets of $S$ and $\caC(S)$ denote the set of all clusters. A function $g:\caP(S) \mapsto \bbC$ is called a \emph{weight} whenever
\begin{enumerate}
\item[i)]
$g(\emptyset) = 1$,
\item[ii)]
If $\De_1 \sim \De_2$, then $g(\De_1 \cup \De_2) = g(\De_1)\,g(\De_2)$.
\end{enumerate}
If the extra condition
\begin{enumerate}
\item[iii)]
$g(\De) = 0$ whenever there is an $X \in \De$ such that $X \nsim \De \setminus \{X\}$
\end{enumerate}
holds true, then we obtain the familiar \emph{polymer model}. In the
sequel we do not assume Condition~iii) to be necessarily true, unless
stated otherwise.

Note a simple duality between the classes of polymer and cluster
models: Any cluster model over the graph $G = (S,\nsim)$ is also a
polymer model over the graph
$G' = (\caC(S),\nsim)$.
The other inclusion is also trivially true. A natural application of
this duality is to the polymer models with a complicated nature of
polymers. Such polymers can often be represented as clusters in a new
cluster model with the polymers being simpler geometric objects.

To any set $A \in \caP(S)$ we assign the \emph{partition function} $Z(A)$ by
\begin{equation}
  Z(A) = \sum_{\De \subset A} g(\De)
\end{equation}
The map between the functions $g$ and $Z$ is actually a bijection and
the last equation can be inverted by means of the M\"obius inversion
formula. In particular, we consider the function
$g^T:\caP(S)\mapsto\bbC$ such that the M\"obius conjugated equations
\begin{equation}
  \log Z(A) = \sum_{\De \subset A} g^T(\De) \qquad
  g^T(\De) = \sum_{A \subset \De} (-1)^{|\De \setminus A|} \log Z(A)
\end{equation}
hold true for all $A \in \caP(S)$ and $\De \in \caP(S)$,
respectively. The function $g^T$ is called a \emph{cluster} weight,
the name being justified by the following simple observation:
\begin{lemma}
For any cluster model, $g^T(\De) = 0$ whenever $\De$ is not a cluster.
\end{lemma}
A familiar result about the polymer model is the exponential decay of
the cluster weight $g^T$ under the assumption on a sufficient
exponential decay of the weight $g$, see~\cite{KoPr,MS00}. We use the
above duality to extend this result to the cluster models,
formulating a new condition that can often be easily checked in
applications.

\begin{proposition}\label{prop: cluster model}
Let positive functions $a,b:S\mapsto \bbR^+$ be given such that
either of the following conditions is satisfied:
\begin{enumerate}
\item
(\textbf{Polymer model})\\
Condition~iii) is fulfilled and\footnote{
   We use the convention $\frac{0}{0} = 0$ here.}
\begin{equation}
  \sup_{X \in S} \frac{1}{a(X)} \sum_{Y \nsim X} e^{(a+b)(Y)} |g(Y)| \leq 1
\end{equation}
\item
(\textbf{Cluster model})\\
There is $z:S\mapsto\bbR^+$ satisfying the condition
\begin{equation}\label{eq: cond. on z}
  \sup_{X \in S} \frac{1}{a(X)} \sum_{Y \nsim X} e^{(2a + b)(Y)} z(Y) \leq 1
\end{equation}
such that $|g(\De)| \leq \prod_{X \in \De} z(X)$ for all $\De \in \caP(S)$.
\end{enumerate}
Then,
\begin{equation}\label{eq: cluster model - result}
  \sup_{X \in S} \frac{1}{a(X)} \sum_{\De \nsim X} e^{\sum_{Y \in \De} b(Y)} |g^T(\De)| \leq 1
\end{equation}
\end{proposition}
\begin{proof}
(1) For the case of the polymer models, see~\cite{KoPr} or better \cite{MS00} for the proof. \\
(2) To prove the statement for a cluster model, we represent it as a polymer model over the graph $(\caC(S),\nsim)$ and make use of the above result. Hence, it is enough to show the inequality
\begin{equation}\label{eq: to prove KP}
  \sum_{\De \in \caC(S) \atop \De \nsim X}
  e^{\sum_{Y \in \De} (a+b)(Y)} |g(\De)| \leq a(X)
\end{equation}
for all $X \in S$. Indeed, then one gets
\begin{equation}
  \sum_{\De \in \caC(S) \atop \De \nsim \De_0}
  e^{\sum_{Y \in \De} (a + b)(Y)} |g(\De)| \leq \sum_{Y \in \De_0} a(Y)
\end{equation}
for all $\De_0 \in \caC(S)$ and the statement about the polymer models yields
\begin{equation}
  \sum_{\De^* \in \caC(\caC(S)) \atop \De^* \nsim \De_0}
  e^{\sum_{\De \in \De^*} \sum_{Y \in \De} b(Y)}
  |g^T(\De^*)| \leq \sum_{Y \in \De_0} a(Y)
\end{equation}
where the sum on the LHS is over all clusters incompatible with $\De_0$ in the
polymer model with the set of polymers $\caC(S)$. Since the weights
$g^T(\De)$ of the clusters in the original cluster model are related to the
cluster weights $g^T(\De^*)$ in the polymer model under consideration as
\begin{equation}
  g^T(\De) = \sum_{\De^*:\ \cup_{\De' \in \De^*}\De' = \De} g^T(\De^*)
\end{equation}
we immediately get
\begin{equation}
  \sum_{\De \nsim X} e^{\sum_{Y \in \De} b(Y)} |g^T(\De)|
  \leq \sum_{\De^* \in \caC(\caC(S)) \atop \De^* \nsim X}
  e^{\sum_{\De \in \De^*} \sum_{Y \in \De} b(Y)} |g^T(\De^*)| \leq a(X)
\end{equation}
which is inequality~\eqref{eq: cluster model - result}.

Using the notation $\hat z(X) := z(X)\, e^{a(X) + b(X)}$ and
\begin{equation}
  \frZ_X(A) = \sum_{\De \in \caC(A) \atop \De \ni X}
  \prod_{Y \in \De} \hat z(Y)
\end{equation}
for any $A \in \caP(S)$ and $X \in A$, inequality~\eqref{eq: to prove KP}
follows from the next two lemmas.
\end{proof}
\begin{lemma}\label{lem: rec. for Y}
The function $\frZ_X(A)$ satisfies the recurrence inequality
\begin{equation}
  \frZ_X(A) \leq \hat z(X)
  \exp \bigl[ \sum_{Y \nsim X \atop Y \in A \setminus \{X\}}
  \frZ_Y(A \setminus \{X\}) \bigr]
\end{equation}
\end{lemma}
\begin{proof}
For any cluster $\De$ we split $\De \setminus \{X\}$ into connected components,
i.e.\ a family of clusters $(\De_j)$, and subsequently write:
\begin{equation}
\begin{split}
  \frZ_X(A) &= \hat z(X)  \sum_{\De \in \caC(A) \atop \De \subset A \setminus \{X\}}
  \prod_{j} \prod_{Y \in \De_j} \hat z(Y)
\\
  &\leq \hat z(X) \sum_{n=0}^{\infty} \frac{1}{n!}
  \sum_{Y_1,\ldots,Y_n \in A \setminus \{X\} \atop \forall j:\ Y_i \nsim X}
  \prod_{j=1}^{n} \sum_{\De_j \subset A \setminus \{X\} \atop \De_j \ni Y_j}
  \prod_{Y \in \De_j} \hat z(Y)
\\
  &= \hat z(X) \sum_{n=0}^{\infty} \frac{1}{n!}
  \bigl[ \sum_{Y \nsim X \atop Y \in \A \setminus \{X\}} \frZ_Y(A \setminus \{X\})
  \bigr]^n
\\
  &= \hat z(X) \exp \bigl[ \sum_{Y \nsim X \atop Y \in A \setminus \{X\}}
  \frZ_Y(A \setminus \{X\}) \bigr]
\end{split}
\end{equation}
\end{proof}
\begin{lemma}
Assume that
\begin{equation}\label{eq: cond. on z II}
  \sum_{Y \nsim X} \hat z(Y) e^{a(Y)} \leq a(X)
\end{equation}
Then
\begin{equation}
  \sum_{Y \nsim X} \frZ_Y(S) \leq a(X)
\end{equation}
\end{lemma}
\begin{proof}
We prove the inequality
\begin{equation}
  \frZ_X(A) \leq \hat z(X) e^{a(X)}
\end{equation}
for all $A \in \caP(S)$ and $X \in A$, by induction in the number of polymers
in the set $A$. Assuming that this bound is satisfied whenever $|A| < n$, we
can estimate $\frZ_X(A)$ for $|A| = n$ by using Lemma~\ref{lem: rec. for Y},
condition~\eqref{eq: cond. on z II}, and the induction hypothesis as follows:
\begin{equation}
  \frZ_X(A) \leq \hat z(X) \exp \bigl[ \sum_{Y \nsim X} \hat z(Y) e^{a(Y)}
  \bigr] \leq \hat z(X) e^{a(X)}
\end{equation}
As the statement is obvious for $|A| = 1$, the lemma is proven.
\end{proof}

\section{Interpolating local limit theorem}\label{sec: LLT}

We present here a simple general result that can be useful in the situations where a full local limit theorem statement is not available due to the lack of detailed control on the dependence among random variables the sum of which is under consideration.
For a detailed explanation of the central and the local limit theorems as well as
the analysis of characteristic functions in the independent case, see e.g.\
\cite{Du}.
Here, under only mild assumptions, we prove an asymptotic upper bound on the probabilities in a regime that interpolates between the ones of the central and the local limit theorem. Namely, we have the following result that is a simple generalization of Lemma 5.3 in \cite{EMN}:
\begin{proposition}\label{prop: LLT}
Let $(X_n)_{n\in\bbN}$ be a sequence of random variables and denote by $\psi_n(t)$ the corresponding characteristic functions, $\psi_n(t) = \bbE\, e^{itX_n}$. If  $(A_n)_{n\in\bbN}$, $(\de_n)_{n\in\bbN}$ and $(\tau_n)_{n\in\bbN}$ are strictly positive sequences of reals satisfying the assumptions
\begin{enumerate}
\item[i)]
$\varlimsup_{n\to\infty} A_n \int_{-\tau_n}^{\tau_n} \id t |\psi_n(t)| \leq 2\pi$
\item[ii)]
There is $k > 1$ such that
$\lim_{n\to\infty} \frac{A_n}{\de_n^{k}\,\tau_n^{k-1}} = 0$
\end{enumerate}
then
\begin{equation}\label{eq: LLT-cond}
  \varlimsup_{n\to\infty} \frac{A_n}{\de_n}\,
  \bsP\{a\de_n \leq X_n \leq b\de_n\} \leq b - a
\end{equation}
for any $a < b$.
\end{proposition}
\begin{remark}
Note that:
\begin{enumerate}
\item
Up to a normalization factor, Condition i) of the proposition only requires $A_n$ to be chosen as
\begin{equation}
  A_n = \caO\Bigl( \Bigl[\int_{-\tau_n}^{\tau_n} \id t\, |\psi_n(t)| \Bigr]^{-1} \Bigr)
\end{equation}
\item
If there is $\ve_1$ such that $A_n \tau_n \leq n^{\ve_1}$ eventually in $n$, then Assumption~ii) of the proposition  is satisfied whenever $\de_n \tau_n \geq n^{\ve_2}$ with a constant $\ve_2 > 0$.
\item
The choice $\de_n = A_n$ (if available) gives an upper-bound on the probabilities in the regime of the central limit theorem. On the other hand, $\de_n = \text{const}$ corresponds to the regime of the local limit theorem. However, for the latter choice
it can be difficult to check the assumptions, and that is why
one has to allow for a sufficient scaling of $\de_n$, see Part~(2) of this remark.
\item
Much more information about the distribution of the random variables $X_n$ would be needed in order to get any \emph{lower bounds} on the probabilities (except for the case $\tau_n = \infty$ in which a full local limit theorem can be proven). This is a hard problem that we do not address here.
\end{enumerate}
\end{remark}
\begin{proof}
Let sequences $(A_n), (\tau_n), (\de_n)$ be given such that the assumption of the proposition is true and take an arbitrary positive function $h \in C^\infty(\bbR)$ for which i) $h(x) = 0$ for any $x \not\in (-\ve,\ve)$ and ii)
$\int_{-1}^{1} \id x\, h(x) = 1$.
Using the notation $G_n$ for the distribution function of $X_n$, we consider its `regularized version' $\bar G_n$ defined by the Lebesgue density
\begin{equation}
  \frac{\id\bar G_n(x)}{\id x} = \int_{-\infty}^{\infty} \id G_n(y)\,
  h_n(x-y)
\end{equation}
where $h_n(x) := \frac{1}{\de_n} h(\frac{x}{\de_n})$. Obviously,
$\frac{\id \bar G_n}{\id x} \in C^\infty(\bbR)$ and it can be expressed by the Fourier integral as follows:
\begin{equation}
\begin{split}
  \frac{\id \bar G_n(x)}{\id x} &= \frac{1}{2\pi}\int_{-\infty}^{\infty} \id t\,
  e^{-itx} \psi_n(t) \int_{-\infty}^{\infty} \id y\, e^{ity} h_n(y)
\\
  &= \frac{1}{2\pi}\int_{-\infty}^{\infty} \id t\,
  e^{-itx} \psi_n(t)\, \hat h(t \de_n)
\end{split}
\end{equation}
where $\hat h(t) := \int_{-\infty}^{\infty} \id x\, e^{itx} h(x)$ and we have used that
$\psi_n(t)\, \hat h(t\de_n) \in L^1(\bbR)$ following from Assumption~i) of the proposition
and from the bounds $|\psi_n(t)|,|\hat h(t)| \leq 1$. Moreover, if $k>1$ is such that
Assumption~ii) holds, then, using the bound $|\hat h(t)| \leq  c |t|^{-k}$ which is true with some constant $c$ for all $t \in \bbR \setminus \{0\}$, we obtain the estimate
\begin{equation}
\begin{split}
  \varlimsup_{n\to\infty} \sup_x A_n \frac{\id G_n(x)}{\id x}
  &\leq \frac{1}{2\pi} \varlimsup_{n\to\infty} \Bigl( A_n \int_{-\tau_n}^{\tau_n} \id t\,
  |\psi_n(t)| + \int_{\bbR \setminus [-\tau_n,\tau_n]} \id t\, |\hat h(t\de_n)|
  \Bigr)
\\
  &\leq 1 + \frac{1}{\pi} \frac{c}{k-1} \varlimsup_{n\to\infty}
  \frac{A_n}{\de_n^{k} \tau_n^{k-1}} = 1
\end{split}
\end{equation}
Finally, by using the inequality
\begin{equation}
\begin{split}
  \bsP\{a\de_n \leq X_n \leq b\de_n\} &= \int_{a\de_n}^{b\de_n} \id G_n(y)
  \int_{-\infty}^{\infty} \id x\, h_n(x-y)
\\
  &\leq \int_{(a-\ve)\de_n}^{(b+\ve)\de_n} \id x\, \int_{a\de_n}^{b\de_n}
  \id G_n(y)\,h_n(x-y)
\\
  &\leq \int_{(a-\ve)\de_n}^{(b+\ve)\de_n} \id \bar G_n(x)
\end{split}
\end{equation}
we get
\begin{equation}
  \varlimsup_{n\to\infty} \frac{A_n}{\de_n}
  \bsP\{a\de_n \leq X_n \leq b\de_n\}
  \leq (b-a+2\ve) \varlimsup_{n\to\infty} \sup_x A_n \frac{\id G_n(x)}{\id x}
  \leq b-a+2\ve
\end{equation}
and the proposition follows by taking the limit $\ve\to 0$.
\end{proof}


\bibliographystyle{plain}

\end{document}